\documentclass[fleqn,usenatbib]{mnras}

\usepackage{newtxtext,newtxmath}

\usepackage[T1]{fontenc}
\usepackage{booktabs}
\usepackage{multicol}
\usepackage{longtable}
\usepackage{supertabular}                                             
\usepackage[table]{xcolor} 
\usepackage{caption}
\usepackage[labelfont=bf]{caption}
\DeclareRobustCommand{\VAN}[3]{#2}
\let\VANthebibliography\thebibliography
\def\thebibliography{\DeclareRobustCommand{\VAN}[3]{##3}\VANthebibliography}

\usepackage{graphicx}	
\usepackage{amsmath}	

\title[The accretion disc in BH-LMXB MAXI\,J1820$+$070]{An empirical connection between line-emitting regions and X-rays heating the accretion disc in BH-LMXB MAXI\,J1820$+$070}

\author[B.E. Tetarenko et al.]{
B.E. Tetarenko$^{1,2}$\thanks{E-mail: bailey.tetarenko@mcgill.ca},
A.W. Shaw$^{3}$, and
P.A. Charles$^{4}$
\\
$^{1}$Department of Physics, McGill University, 3600 University Street, Montr\'{e}al, QC H3A 2T8, Canada\\
$^{2}$Trottier Space Institute at McGill, McGill University, 3550 University Street, Montr\'{e}al, QC H3A 2A7, Canada\\
$^{3}$Department of Physics, University of Nevada, Reno, NV 89557, USA\\
$^{4}$Department of Physics \& Astronomy, University of Southampton, Southampton SO17 1BJ, UK\\
}

\date{Accepted XXX. Received YYY; in original form ZZZ}

\pubyear{2023}

\begin{document}

\label{firstpage}
\pagerange{\pageref{firstpage}--\pageref{lastpage}}
\maketitle

\begin{abstract}
The recurring transient outbursts in low-mass X-ray binaries (LMXBs) provide ideal laboratories to study the accretion process. Unlike their supermassive relatives, LMXBs are far too small and distant to be imaged directly. Fortunately, phase-resolved spectroscopy can provide an alternative diagnostic to study their highly complex, time-dependent accretion discs. The primary spectral signature of LMXBs are strong, disc-formed emission lines detected at optical wavelengths. The shape, profile, and appearance/disappearance of these lines change throughout a binary orbit, and thus, can be used to trace how matter in these discs behaves and evolves over time. By combining a \textit{Swift} multi-wavelength monitoring campaign, phase-resolved spectroscopy from the Gran Telescopio Canarias (GTC) and Liverpool Telescope, and modern astrotomography techniques, we find a clear empirical connection between the line emitting regions and physical properties of the X-rays heating the disc in the black hole LMXB MAXI J1820+070 during its 2018 outburst. In this paper, we show how these empirical correlations can be used as an effective observational tool for understanding the geometry and structure of a LMXB accretion disc and present further evidence for an irradiation-driven warped accretion disc present in this system.

\end{abstract}




\section{Introduction}

Low-mass X-ray binaries (LMXBs) and Cataclysmic Variables (CVs) are the ideal laboratories to study both the inflow of matter through, and the outflows driven from, astrophysical accretion discs. Their accretion discs, which feed compact stellar remnants (i.e., black holes, neutron stars, and white dwarfs) and are fed by nearby low-mass stars ($\lesssim1M_{\odot}$), undergo recurrent (accretion-driven) outbursts. 
These outbursts, which typically take place on time-scales of days-months, provide a unique opportunity to study an evolving accretion disc in real-time \citep{charlescoe2006, remillardmclintock2006,tetarenko2016}.

While impossible to image directly, phase-resolved spectroscopy can provide an indirect view of their highly complex, time-dependent accretion discs. Strong disc-formed recombination emission lines (Balmer series, He {\sc i}, He {\sc ii}) are the primary spectral feature of LMXBs at optical wavelengths \citep{charlescoe2006}. These lines often display double-peaked profiles, due to Doppler motions within the binary \citep{crawfordkraft1956,casares2015}, with shape depending on the distribution of emission over the disc surface \citep{marsh2001,marsh2005}. The strength, profile, and appearance and disappearance of these lines change throughout a binary orbit, and thus can be used to trace how accretion disc gas behaves and evolves over time \citep{marsh2001,charlescoe2006}.

LMXBs harbouring stellar-mass black holes (BHs; $5-15M_{\odot}$) are of particular interest.
Most of the optical light emitted by their discs comes from the outer disc regions, which reprocess X-rays produced close to the BH. By illuminating the disc surface, this X-ray irradiation is the dominant factor determining temperature over most of the disc during outburst. Additionally, it controls overall outburst evolution from peak to quiescence and determines the amount of mass accreted and lost via the disc itself \citep{vanparadijs1994,vanparadijs1996,charlescoe2006}.

For such emission lines to be produced in this environment, a temperature inversion must exist in the disc atmosphere \citep{shaviv1986,hubeny1990}. Thus, X-ray irradiation has long been thought the likely process behind line production in LMXBs. As irradiation is most significant in the cool, outer disc ($\gtrsim$ hundreds of gravitational radii), disc-formed H/He emission lines can be used as powerful diagnostic tracers of its source and effect on the structure and geometry of the gas making up the accretion disc (e.g., \citealp{tetarenko2021}).

The Galactic BH-LMXB MAXI\,J1820$+$070 (hereafter J1820), was initially discovered by the All-sky Automated Survey for Supernovae (ASAS-SN) as an optical transient (ASASSN-18ey; \citealp{tucker2018}), and later first detected at X-ray wavelengths \citep{kawamaro2018} with the Monitor of All-sky Image (MAXI), in March of 2018. J1820 was soon classified as a candidate BH-LMXB based on its multi-wavelength behaviour (e.g., \citealp{kennea2018, baglio2018,bright2018,shidatsu2018,russell2019a,russell2019b}) and later dynamically confirmed to contain a stellar-mass BH once in quiescence \citep{torres2019}.
J1820 would remain active for multiple months, subsequently becoming one of the brightest (at both optical and X-ray wavelengths) known BH-LMXBs ever observed \citep{shidatsu2019,russell2019a}, resulting in the system being extensively observed by many ground and space based observatories.

During this 2018 outburst, J1820 underwent the hysteretic behaviour typical of BH-LMXBs (see e.g., \citealt{tetarenko2016}), first detected during the outburst rise in the hard state, followed by a transition to the soft state \citep{shidatsu2019,fabian2020}, during which a relativistic jet outflow was observed \citep{homan2020,bright2020}. Additionally, the optical outburst spectrum showed multiple broad emission lines \citep{tucker2018b}, characteristic of outbursting LMXBs, as well as signatures for the presence of an accretion disc wind outflow \citep{munozdarias2019}.
Lastly, through comprehensive photometric monitoring, large modulations in the optical light-curve were also observed, evolving from the superhump period \citep{patterson2018} to close to the true orbital period \citep{torres2019} of the system throughout outburst. Detailed temporal analysis has since interpreted this behaviour in the context of an evolving warped accretion disc present in the system \citep{thomasJ2022}.

In this paper, we use a combination of phased-resolved optical spectroscopy, quasi-simultaneous with X-ray spectroscopy and multi-wavelength photometric monitoring, to build empirical correlations between the disc-formed emission line profiles and physical properties of the X-rays heating the disc, in J1820. Correlations built from multiple lines allow us to probe different regions of J1820's accretion disc, and ultimately prove to be a useful observational tool for understanding the evolution of the geometry and structure of an outbursting LMXB accretion disc.

\section{Observations and Data Reduction}\label{sec:obs_details}

\subsection{Liverpool Telescope (LT)}\label{sec:LT_data}

Between 2018 March 17 and 2018 June 18, optical spectra of J1820 were obtained using the Fibre-fed RObotic Dual-beam Optical Spectrograph (FRODOspec; \citealp{MoralesRueda2004}), an integral field unit spectrograph on the 2m Liverpool Telescope located at the Roque de los Muchachos Observatory in La Palma, Spain. A total of 175 epochs with 600-900s exposure times were taken over 38 nights, using a combination of the VPH gratings on the red (5900-8000\AA) and blue (3900-5100\AA) arms (see Tables \ref{tab:frodoblue} and \ref{tab:frodored}). 
Data were reduced in three steps. First, raw data were run through the CCD processing pipeline (L1; \citealp{barnsley2012}), which performs bias subtraction, overscan trimming, and CCD flat fielding procedures. Second, cosmic-ray rejection was performed on the L1 pipeline products using the LA-Cosmic Program \citep{lacosmicpaper} in {\sc pyraf 2.1.15}. Lastly, the cleaned data were run through the FRODOspec reduction pipeline\footnote{\url{https://github.com/LivTel/frodo-l2-pipeline}} (L2; see \citealp{barnsley2012} for details), which produces a reduced, extracted, throughput-corrected and wavelength-calibrated spectrum.

\subsection{Gran Telescopio Canarias (GTC)}\label{sec:GTC_data}

Between 2018 March 17 and 2018 November 21, optical spectra of J1820 were obtained using the Optical System for Imaging and low Resolution Integrated Spectroscopy (OSIRIS; \citealp{cepa2000}) instrument, an imager and spectrograph on the 10.4m GTC, located at the Roque de los Muchachos Observatory in La Palma, Spain. A total of 93 spectral epochs were taken over 22 nights with varying exposure times, using a 1\arcsec slit and both the R2500V (4500-6000\AA) and R2500R (5575-7684\AA) grisms (see Tables \ref{tab:gtcV} and \ref{tab:gtcR}). Standard procedures in {\sc iraf} were used to reduce, extract, and wavelength and flux calibrate the long-slit spectral data. See \citet{munozdarias2019} for details on the reduction.

\subsection{Swift}\label{sec:swift_data}

\begin{figure*}
    \center
    \includegraphics[width=1.0\linewidth,height=0.95\linewidth]{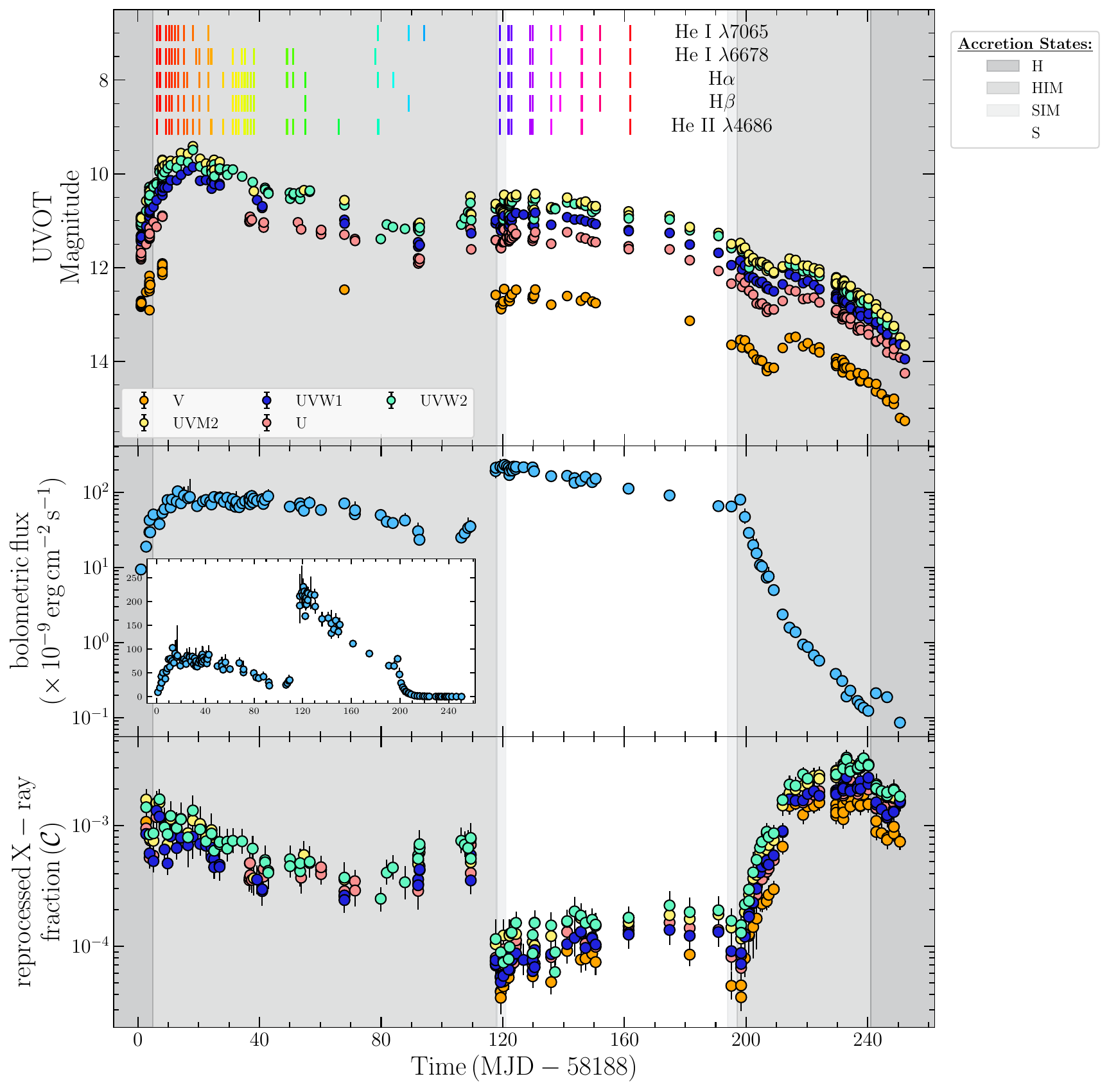}
    \caption{The evolution of the 2018 outburst of BH-LMXB J1820: \textit{(top)} the de-reddened UV and optical light-curve as observed by \textit{Swift}/UVOT, \textit{(middle)} the bolometric light-curve (computed from $0.5-10$ keV \textit{Swift}/XRT data; see Section \ref{sec:accstate}) in log and linear space, and \textit{(bottom)} the fraction of X-rays intercepted and reprocessed in the outer disc ($\mathcal{C}$) as a function of time (calculated with the \citealp{tetarenko2020} method and the five UVOT filters available; see Section \ref{sec:xrayirr}). All uncertainties are quoted to the 1$\sigma$ confidence level.
Shaded background (grey) regions show the accretion state evolution of the source over time (see Section \ref{sec:accstate} and the legend for details). Coloured lines in the top panel display times where there exist epochs of optical spectroscopy (GTC/OSIRIS and LT/FRODOspec) quasi-simultaneous (within $<1$d) with the \textit{Swift} monitoring observations.}
    \label{fig:lc_plot}
\end{figure*}

Between 2018 March 12 and 2018 November 16, a total of 128 observations, taken with instruments aboard the Neil Gehrels Swift Observatory, were obtained from the High Energy Astrophysics Science Archive Research Center (HEASARC) Archive\footnote{\url{https://heasarc.gsfc.nasa.gov/docs/archive.html}}, covering the 2018 outburst of J1820. Observations using the X-ray Telescope (XRT; \citealp{burrows2005}) were largely taken in windowed timing (WT) mode, with the exception of a few instances when the source count rate was low (count rate $<$0.7 counts/s) and photon counting (PC) mode was appropriate. Observations using the UV and Optical Telescope (UVOT; \citealp{roming2005}) were obtained in the  UVW2, UVM2, UVW1, U and V filters. See Figure \ref{fig:lc_plot} for the 2018 outburst light-curve of J1820.

\subsubsection{XRT}

Using the {\sc heasoft} v6.30.1 software package, XRT data was initially processed using the {\tt xrtpipeline} task. Next, source and background spectra were extracted. For all WT mode observations, 20\arcsec radii circular apertures were used. For PC mode observations with an average count rate $<$0.5 counts/s, the same procedure as WT mode was used. If PC mode observations had an average count rate $>$0.5 counts/s, spectra were extracted using standard procedures used to 

\noindent deal with pile-up\footnote{\url{https://www.swift.ac.uk/analysis/xrt/pileup.php}}. For the source spectra, an annular region with a 20 pixel outer radius, and an excluded inner region whose size was calculated via the {\tt ximage} tool, was used. For the background spectra, an annulus with inner
and outer radii of 50 and 70 pixels, respectively centered on the source, was used. Following spectral extraction, the {\tt grppha} task was used to group source and background spectra to have at least 15 counts per energy bin. Finally, the HEASARC calibration
data base (CALDB) and the {\tt xrtmkarf} task were used to obtain/generate the response matrix and ancillary response files, respectively.

\subsubsection{UVOT}

Aperture photometry was performed, on all available UVOT observations taken simultaneous
with the XRT exposures described above, using the {\sc heasoft} software task {\tt uvotsource}. Regions with radii of 5\arcsec (centred on the source) and 20\arcsec (in a source
free region) were used for the source and background, respectively. UVOT magnitudes were computed in the Vega system and uncertainties on these magnitudes account for both statistical error and error in the shape of the Point Spread Function (PSF). All uncertainties are quoted to the 1$\sigma$ confidence level. UVOT data were corrected for interstellar extinction using \citet{fitzpatrick1999}, and subsequently de-reddened using an $E(B-V)=0.18$ \citep{tucker2018}.

\section{Analysis and Results}

\subsection{X-ray Spectral Fitting}\label{sec:xrayspecfit}

Spectral fitting was performed in {\sc xspec} v12.12.1 in the 0.5--10keV band. All X-ray spectra were adequately fit (using $\chi^2$ statistics) with one of the following three models, an: absorbed power-law ({\tt tbabs*powerlaw}), absorbed disc-blackbody ({\tt tbabs*diskbb}), or power-law + disc blackbody ({\tt tbabs*(diskbb+powerlaw})). Abundances from \citet{wilms2000}, and photoionization cross-sections from \citet{verner1996}, were utilized within the {\tt tbabs} model to account for interstellar absorption in all fits. Band-limited fluxes were computed from the best-fit models to the X-ray spectra.

\subsection{Accretion State Classification}\label{sec:accstate}

Using the classification scheme defined in \citet{marcel2019}, we classify all X-ray observations of J1820 into one of four accretion states: hard (H), hard-intermediate (HIM), soft-intermediate (SIM) and soft (S). See Figure \ref{fig:lc_plot} for details. This method defines accretion state based on the: (i) power-law fraction (ratio of the power-law flux to the total flux), and (ii) photon index, computed using the best-fits to the XRT spectra (see Section \ref{sec:xrayspecfit}). Bolometric fluxes were computed using standard bolometric corrections valid for LMXBs, estimated for each accretion state by \citet{migliari2006}.

\subsection{Evolution of the X-ray Irradiating Source}\label{sec:xrayirr}

Applying the computational technique developed by \citet{tetarenko2020} to the Swift data (XRT and UVOT; see Section \ref{sec:swift_data}) available, we have been able to track the evolution of the X-ray source heating the disc in J1820 during the 2018 outburst (see Figure \ref{fig:lc_plot}). This technique statistically compares the X-ray (proxy for the rate matter moves through the disc and falls into the BH) against the UV/Optical (only direct probe of X-ray heated disc gas) emission, to compute how the fraction of X-rays intercepted and reprocessed in the outer disc ($\cal C$) varies throughout an outburst cycle.

The following binary orbital parameters were assumed for this analysis: orbital period $P_{\rm orb}=0.68549\pm0.00001$d, mass ratio $q=0.072\pm0.012$, BH mass $M_1=8.48^{+0.79}_{-0.72}M_{\odot}$, inclination $i=63^{\circ}\pm3$, and distance of $2.96\pm0.33$kpc \citep{torres2019,torres2020,atri2020}.

\subsection{Emission Line Analysis}\label{sec:emissionlines}

\begin{figure*}
    \center
    \includegraphics[width=0.7\linewidth,height=1.25\linewidth]{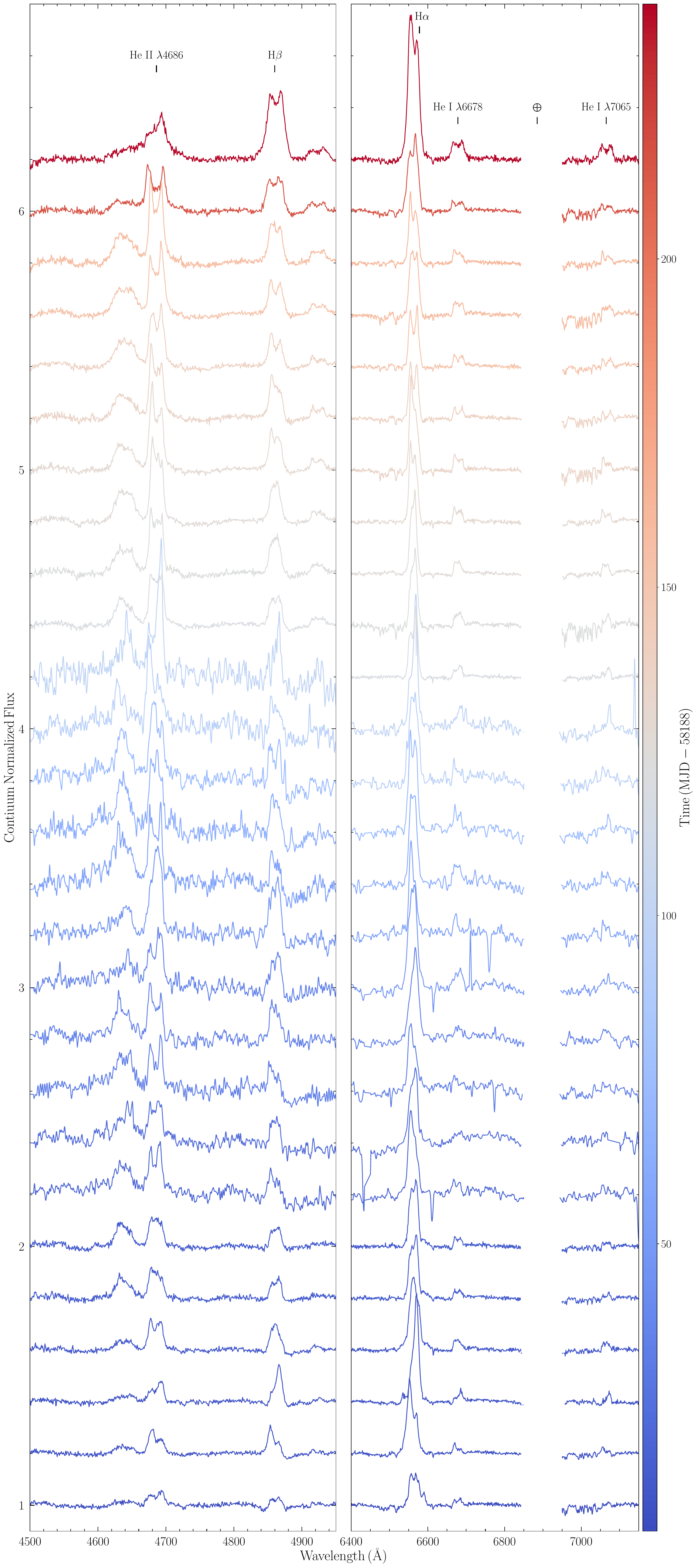}
    \caption{Sample optical spectra, taken with the LT/FRODOspec and GTC/OSIRIS, showing the evolution of the line profiles (for the five emission lines of interest: H$\alpha$, H$\beta$, He {\sc ii} $\lambda4686$, He {\sc i} $\lambda6678$, and He {\sc i} $\lambda7065$) as the source evolves through accretion states during the 2018 outburst of J1820: initial rise/decay H/HIM (dark to light blue), SIM-S (grey to light red), final decay H/HIM (dark red). Tellurics (marked with $\bigoplus$) have been removed for readability. From bottom to top, spectra 7-17 were taken with LT, while spectra 1-6 and 18-27 were taken with GTC. Note that a handful of sharp (unreal) emission and absorption features appear in some spectra from the red arm of LT/FRODOspec. These are likely caused by cosmic ray and sky subtraction issues. Neither has any effect on our actual analysis. }
    \label{fig:spec_examples}
\end{figure*}

The optical spectrum of J1820 contains a number of strong double-peaked H and He emission lines, typical of an LMXB accretion disc. We focus our analysis on the five emission lines present in both GTC and LT spectra: H$\alpha$, H$\beta$, He {\sc ii} $\lambda4686$, He {\sc i} $\lambda6678$, and He {\sc i} $\lambda7065$. See Figure \ref{fig:spec_examples} for a set of sample spectra taken throughout outburst.

We fit a double Gaussian model to these five emission lines of interest 
using an MCMC algorithm that is implemented in {\sc python} using the {\tt emcee} package \citep{foremanmackay2013}. We use the {\tt pyHarmonySearch} \citep{geem2001} global optimization algorithm, which provides a brute-force grid search of the parameter space, to compute a starting point for the MCMC algorithm. We set Gaussian priors for each of the 6 model parameters, with the mean of the distribution set using the {\tt pyHarmonySearch} result.

After initialization, the MCMC algorithm is run on each continuum-normalized spectral epoch, cut around each of the five emission lines (3000km/s on either side of the line center), using a number of walkers equal to 10 times the model dimensions. This process involves a 500 step ``burn-in'' phase (allowing the walkers to sufficiently explore the parameter space), followed by a 1000 step ``sampling'' phase (whereby the MCMC sampler is run again until convergence). We take the median and 1$\sigma$ confidence interval of the posterior distributions, built with the MCMC sampler, as the best-fit model for each line. Lastly, for each best-fit model, we compute equivalent width (EW)\footnote{Note that, continuum fitting was done using the entire available spectrum, excluding strong emission/absorption features.} and full-width half max (FWHM) of the emission lines using the { \tt specutils} package in {\sc python}. Uncertainties on EW and FWHM are propagated from the MCMC algorithm results using Monte-Carlo sampling.

\subsection{Empirical Correlations}\label{sec:correlations}

Considering all optical spectral epochs which have corresponding quasi-simultaneous (within $<1$d) Swift monitoring data, we build correlations between emission line profile shape (defined by the EW and FWHM; see Section \ref{sec:emissionlines}) and the evolving strength of high-energy X-rays illuminating the disc ($\mathcal{C}$; see Section \ref{sec:xrayirr} and Figure \ref{fig:lc_plot}), throughout the 2018 outburst of J1820.

\begin{figure*}
    \center
    \includegraphics[width=0.49\linewidth,height=0.49\linewidth]{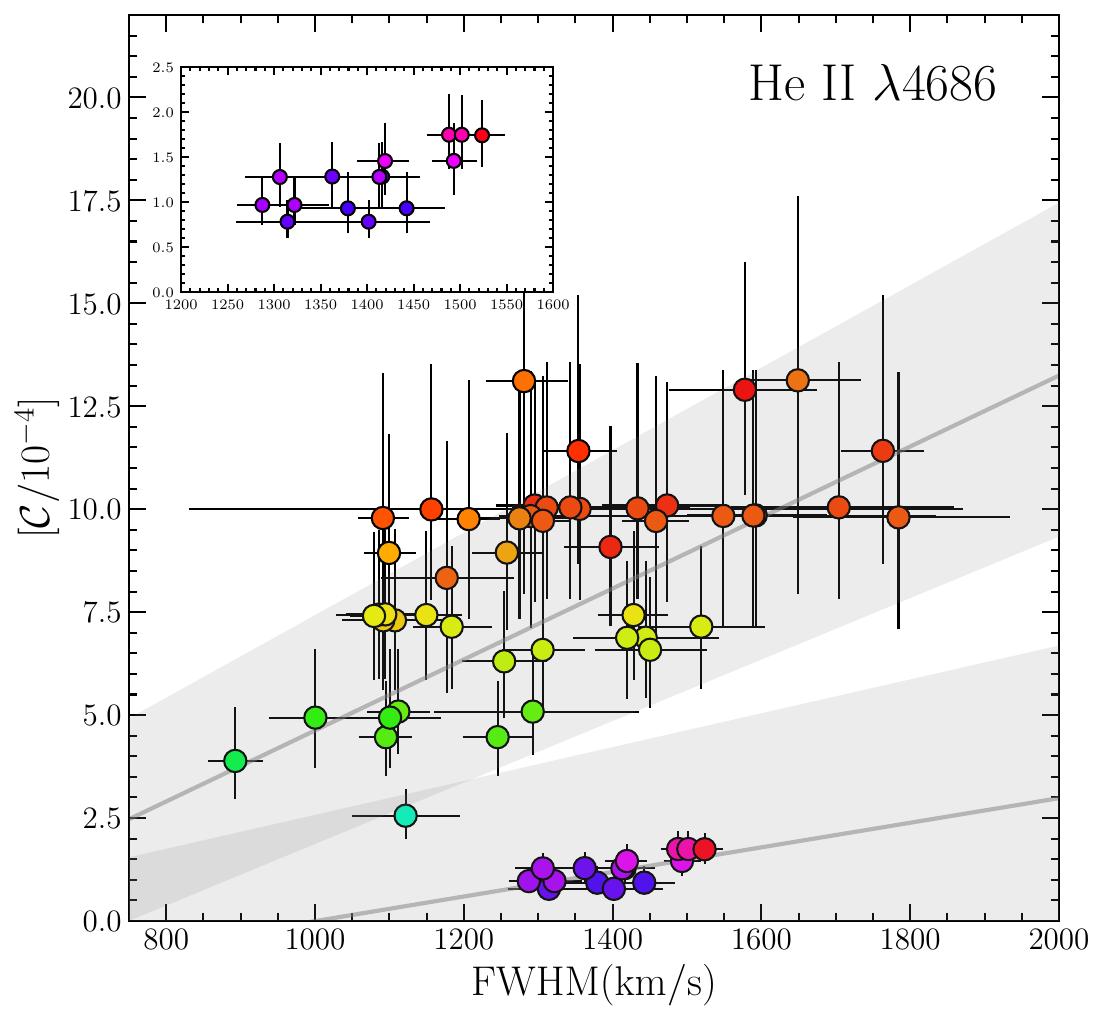}\hfill
    \includegraphics[width=0.49\linewidth,height=0.49\linewidth]{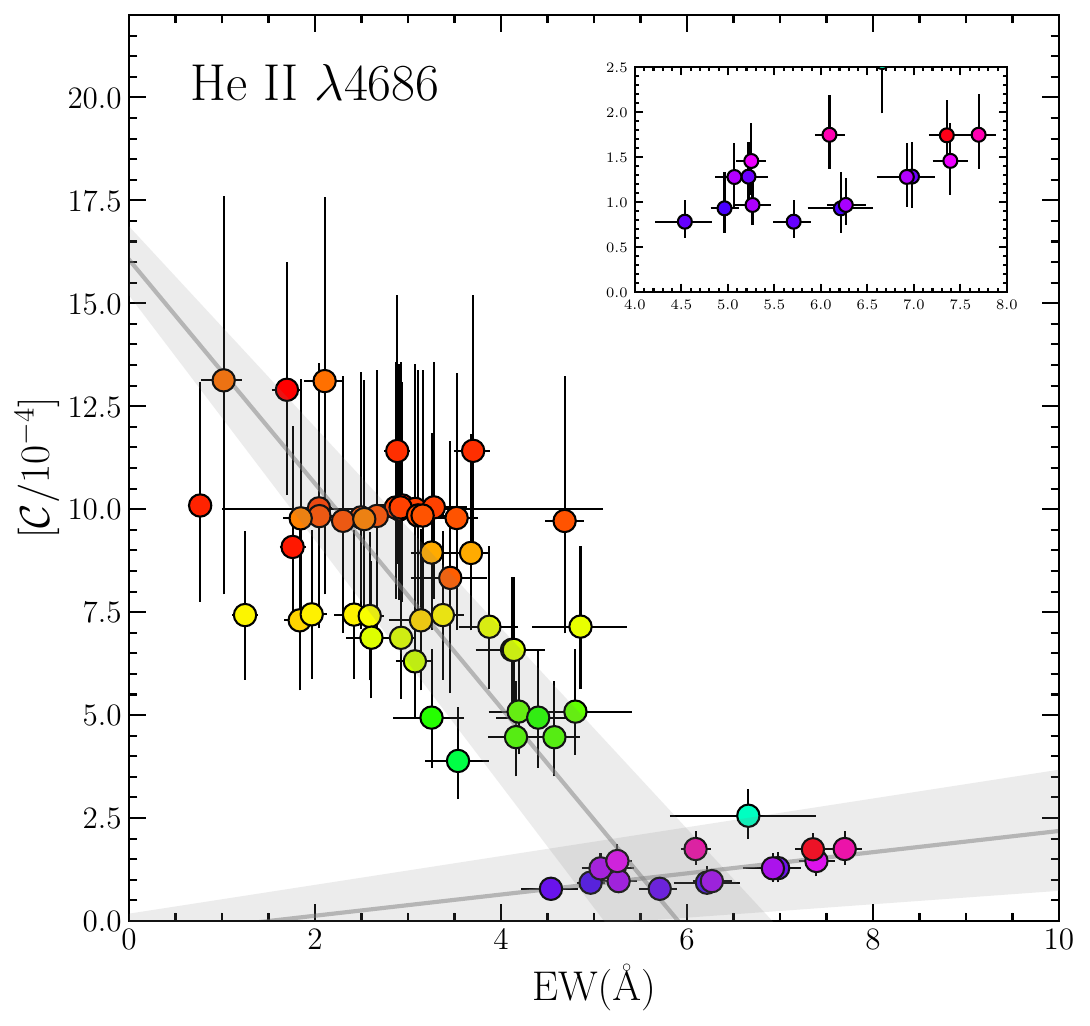}
    \includegraphics[width=0.49\linewidth,height=0.49\linewidth]{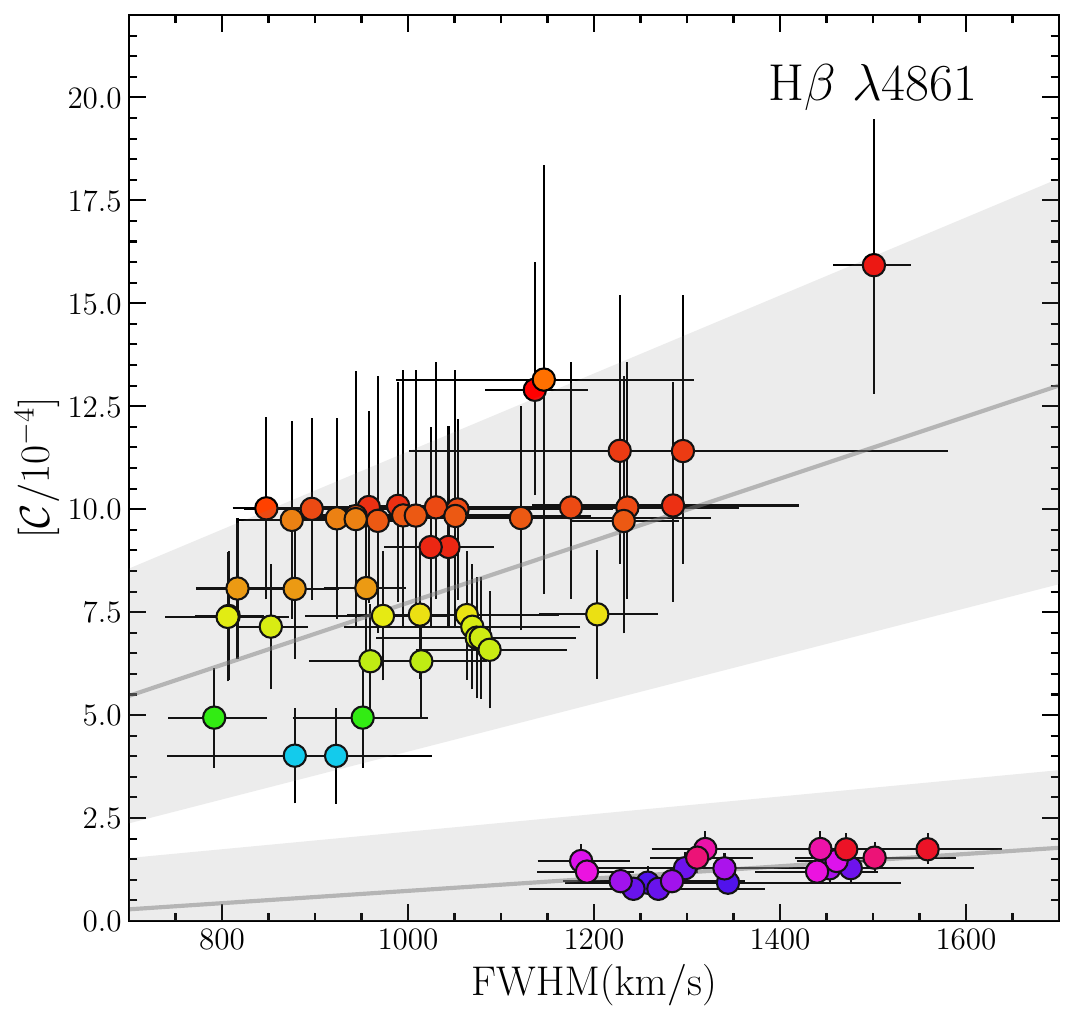}\hfill
    \includegraphics[width=0.49\linewidth,height=0.49\linewidth]{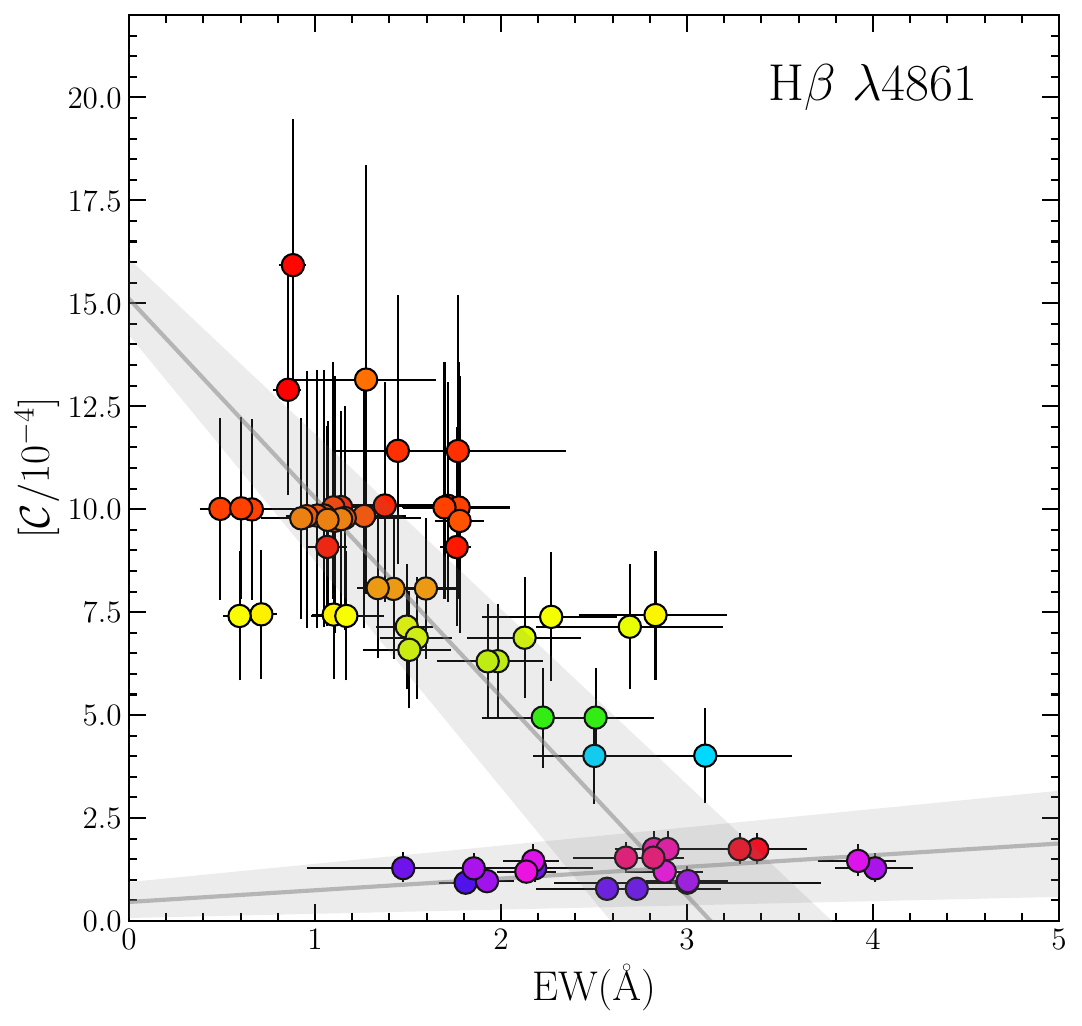}
    \caption{Changes in profile shape, defined in terms of FWHM (\textit{left}) and EW (\textit{right}), of the disc-formed H/He emission lines (as observed with GTC/OSIRIS and LT/FRODOspec), correlated with the evolving strength of high energy X-rays illuminating the disc ($\mathcal{C}$), during the 2018 outburst of J1820. $\mathcal{C}$, defined as the fraction of total X-rays intercepted and reprocessed in the disc, is computed utilizing the computational technique from \citet{tetarenko2020} with the available \textit{Swift} XRT and UVOT/UVW2 monitoring data (see Section \ref{sec:xrayirr}). The UVW2 filter is used as it provides the most complete temporal coverage of the outburst. Only spectral epochs which have corresponding quasi-simultaneous (within $<1$d) Swift monitoring data are used in the correlations. The colour of the data points is representative of the temporal evolution of the outburst (see Figure \ref{fig:lc_plot}). Inset axes are used to show example zoomed-in views of soft state behaviour. Hard and soft accretion state correlations are fit separately with a linear MCMC model (see Section \ref{sec:correlations} and Table \ref{tab:corr_fits}). The best-fit (solid grey lines) and 1$\sigma$ confidence interval of the fits (shaded grey regions) are shown for each correlation.}
    \label{fig:corrs1}
\end{figure*}

\begin{figure*}
    \center
    \includegraphics[width=0.49\linewidth,height=0.49\linewidth]{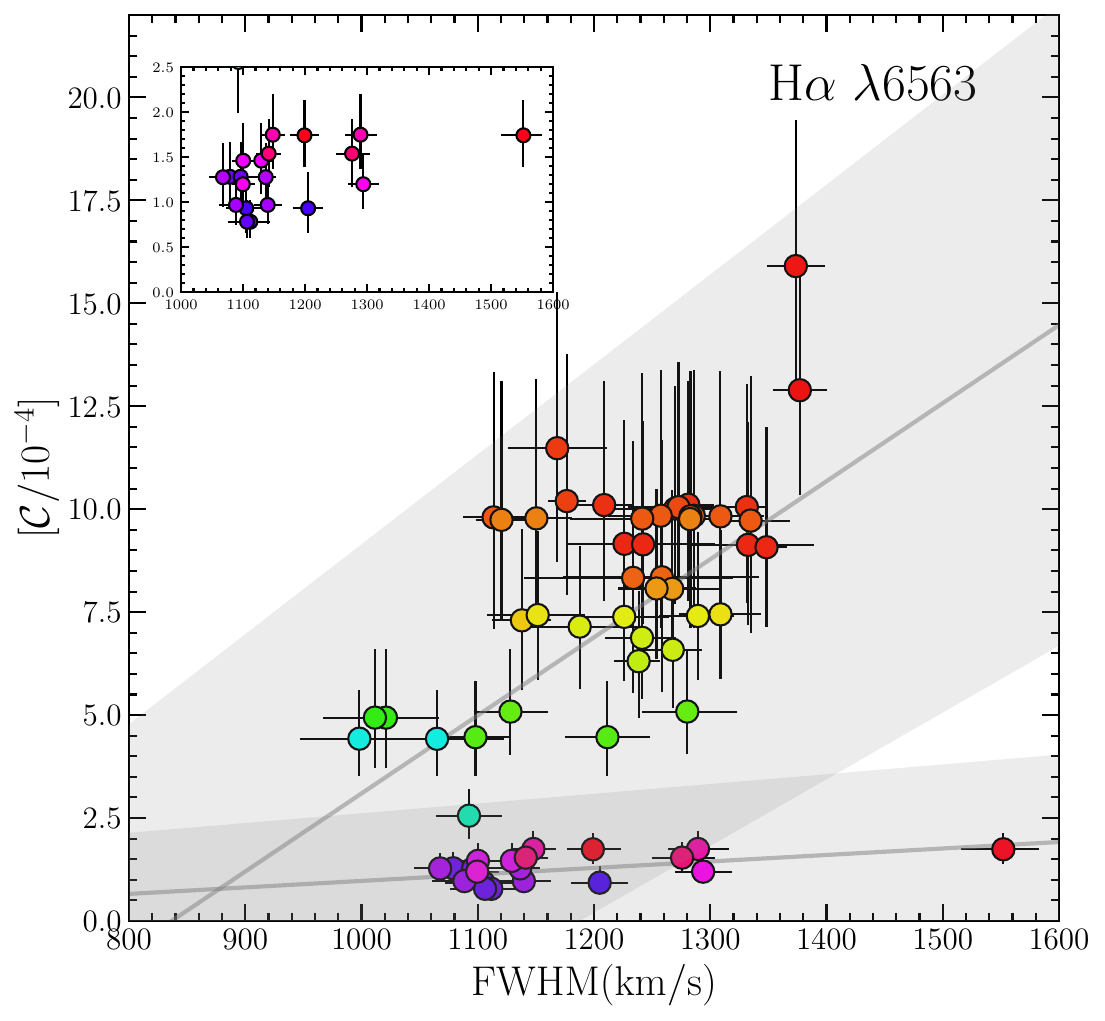}\hfill
    \includegraphics[width=0.49\linewidth,height=0.49\linewidth]{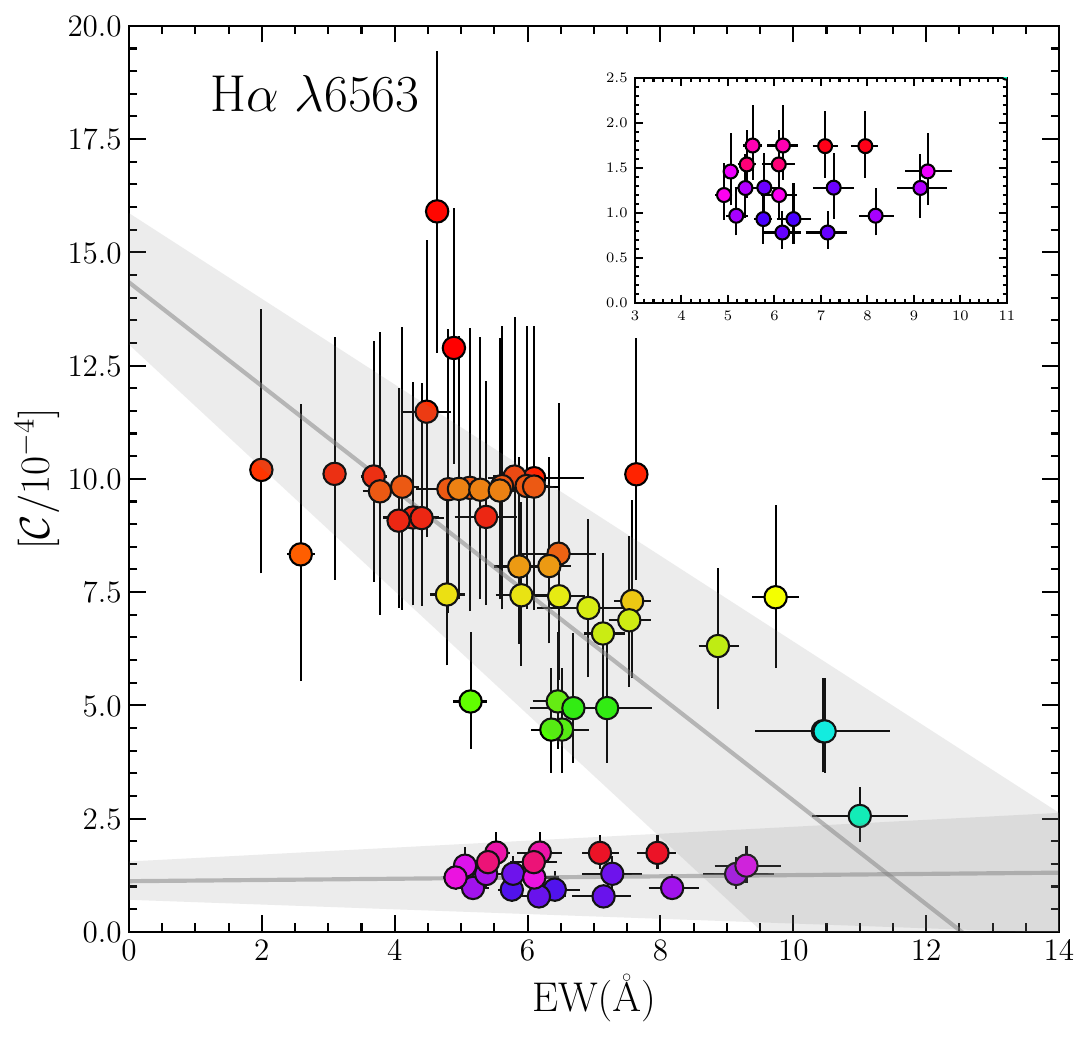}
    \includegraphics[width=0.49\linewidth,height=0.49\linewidth]{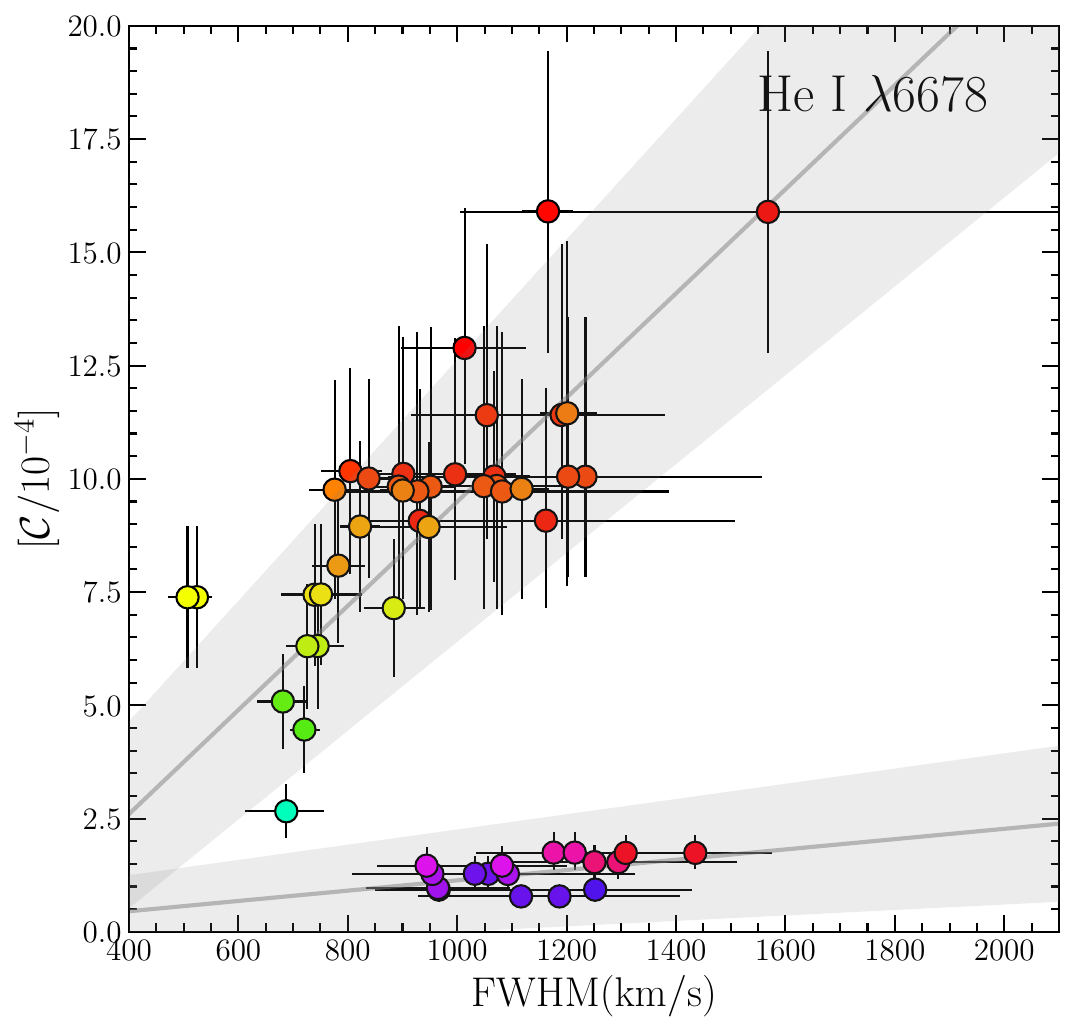}\hfill
    \includegraphics[width=0.49\linewidth,height=0.49\linewidth]{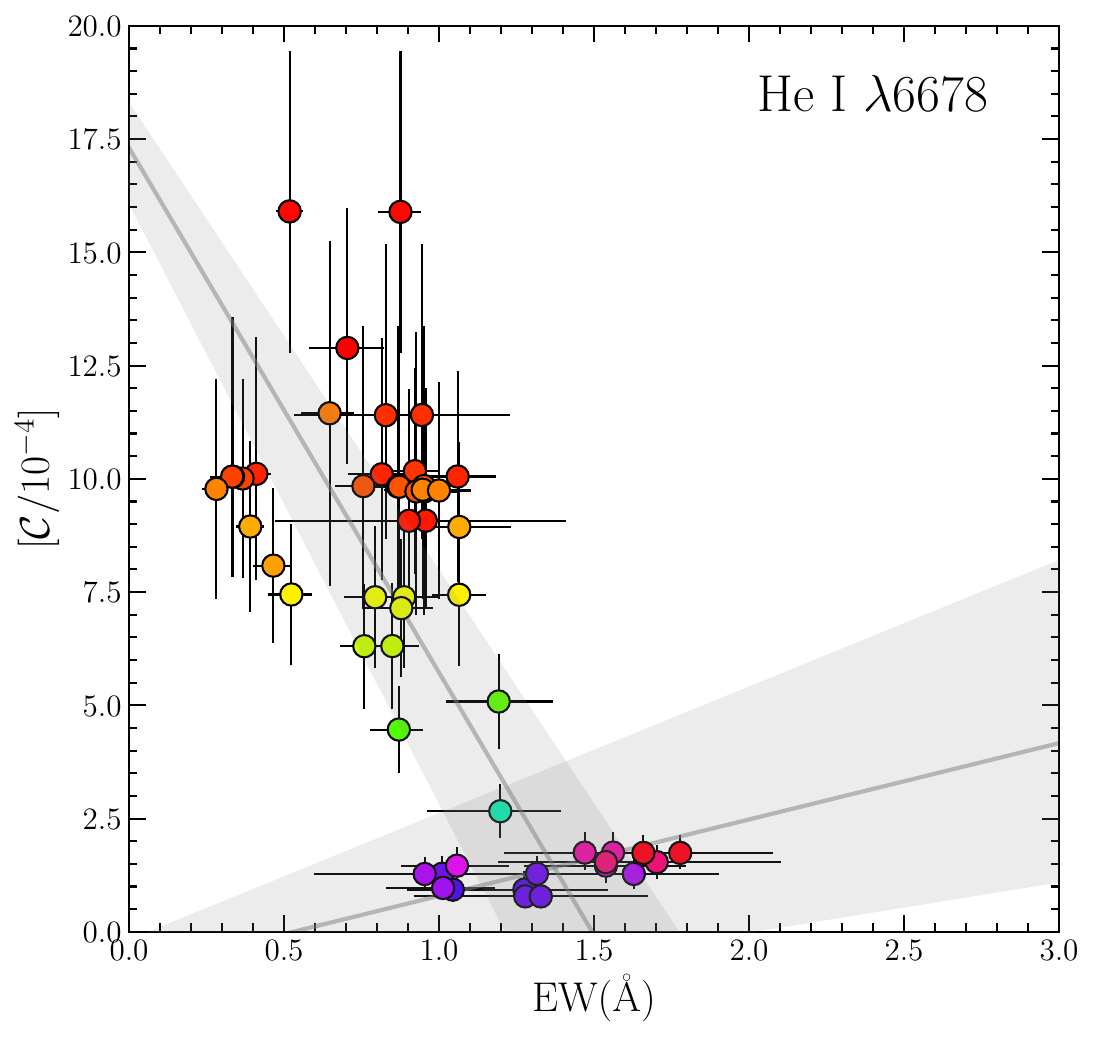}
    \caption{Same as Figure \ref{fig:corrs1} but for H$\alpha$ and He {\sc i} $\lambda6678$.}
    \label{fig:corrs2}
\end{figure*}

\begin{figure*}
    \center
    \includegraphics[width=0.49\linewidth,height=0.49\linewidth]{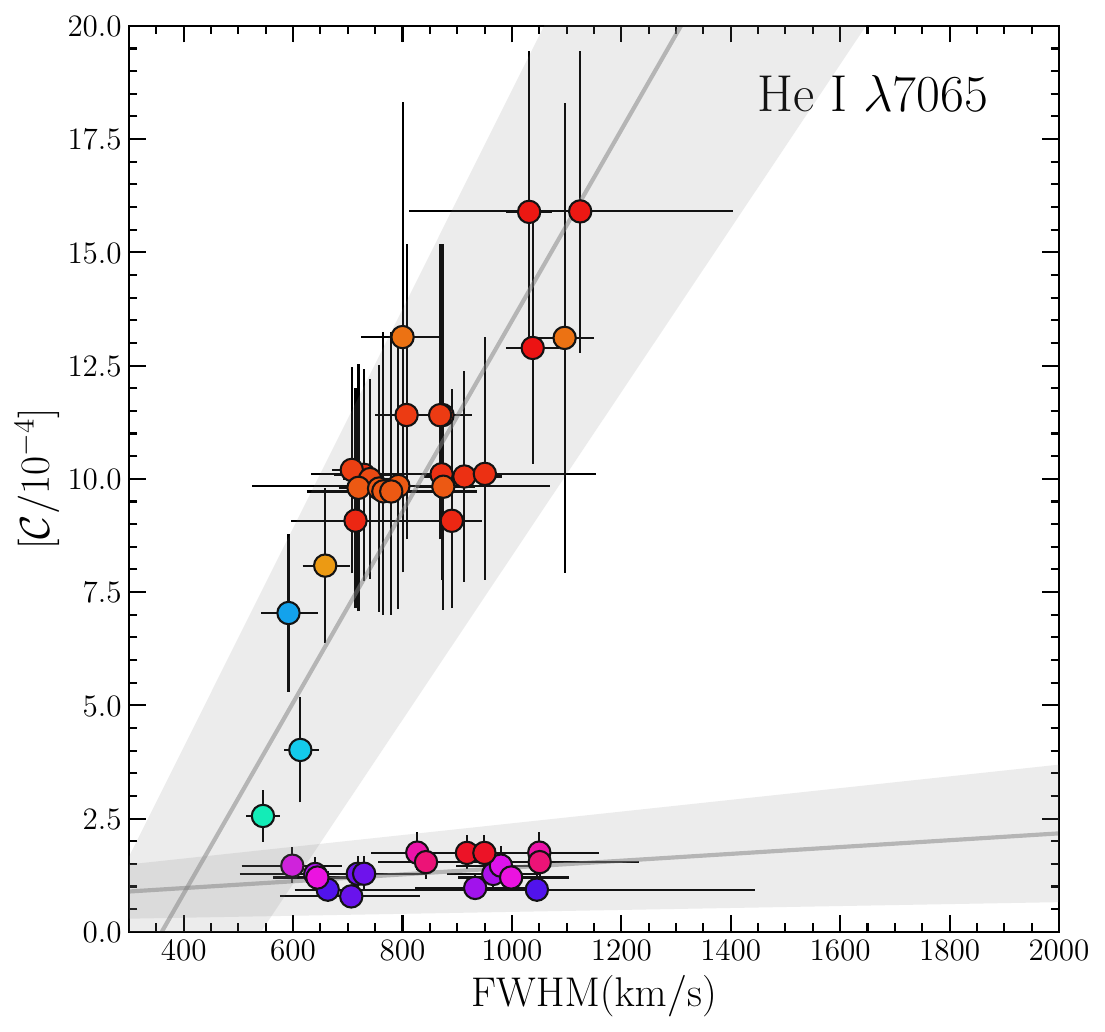}\hfill
    \includegraphics[width=0.49\linewidth,height=0.49\linewidth]{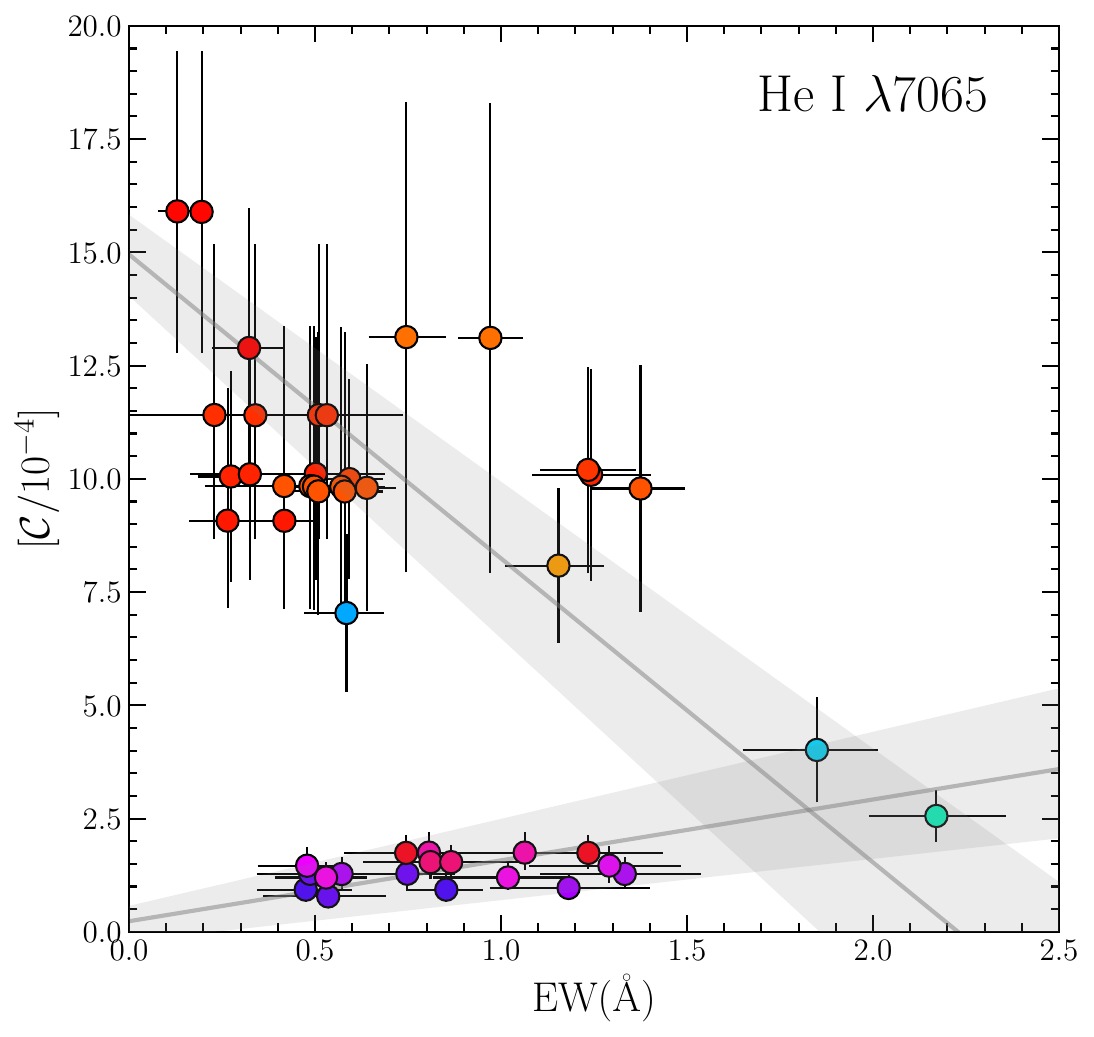}
    \caption{Same as Figure \ref{fig:corrs1} but for He {\sc i} $\lambda7065$.}
    \label{fig:corrs3}
\end{figure*}

We split each correlation by accretion state, grouping H and HIM, and S and SIM, state data together and then fitting the data groups with a linear model using the MCMC algorithm presented in Section \ref{sec:emissionlines}. Correlations between (i) EW and $\mathcal{C}$, and (ii) FWHM and $\mathcal{C}$, for the H$\alpha$, H$\beta$, He {\sc ii} $\lambda4686$, He {\sc i} $\lambda6678$, and He {\sc i} $\lambda7065$ lines, are shown in Figures \ref{fig:corrs1}-\ref{fig:corrs3}. Correlations using $\mathcal{C}$ computed with the UVW2 filter are shown as this filter provides the best outburst overage of all available UVOT filters. Best fit results can be found in Table \ref{tab:corr_fits}.

\begin{figure*}
    \center
    \includegraphics[width=0.34\linewidth,height=0.31\linewidth]{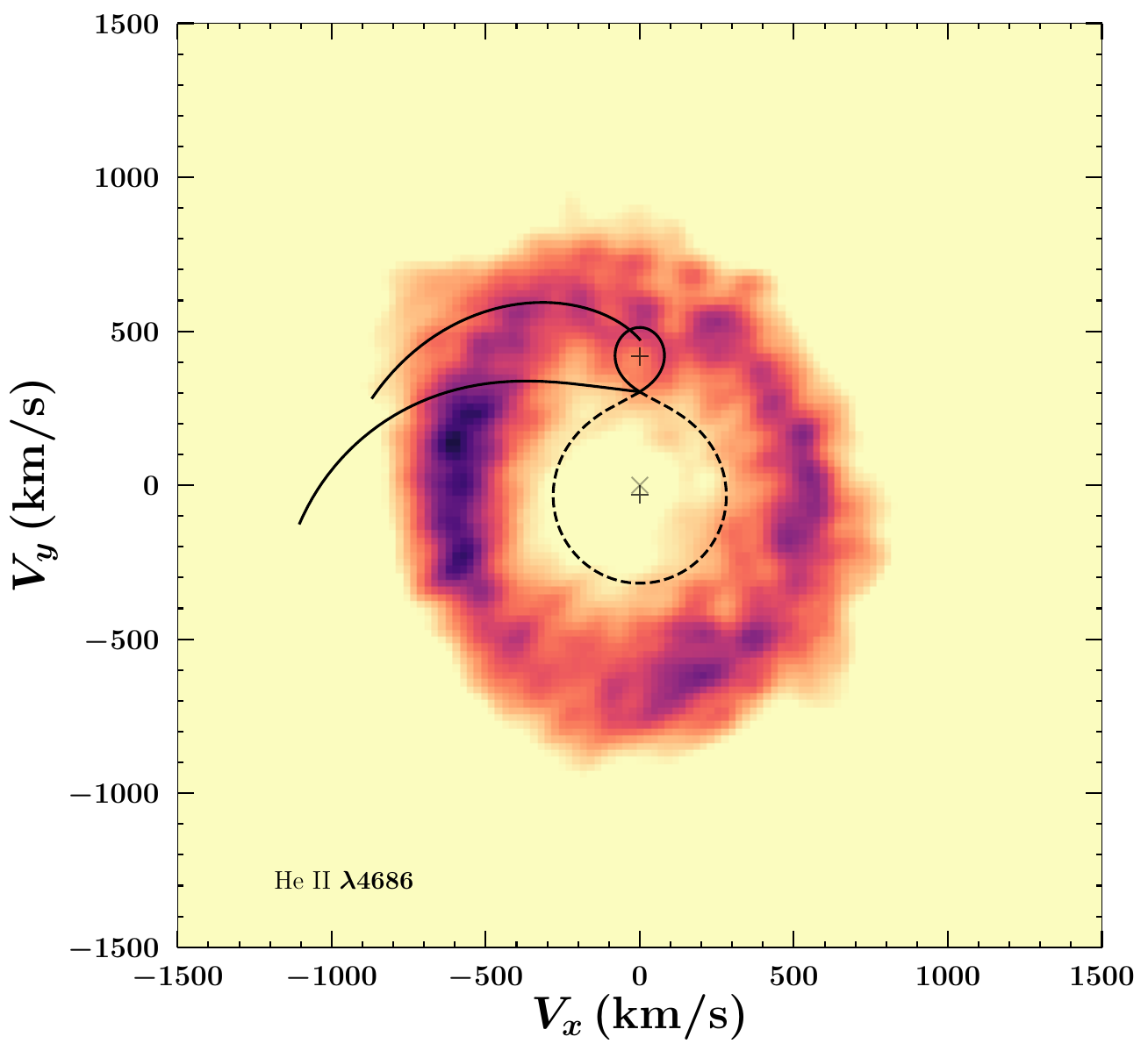}
    \includegraphics[width=0.34\linewidth,height=0.31\linewidth]{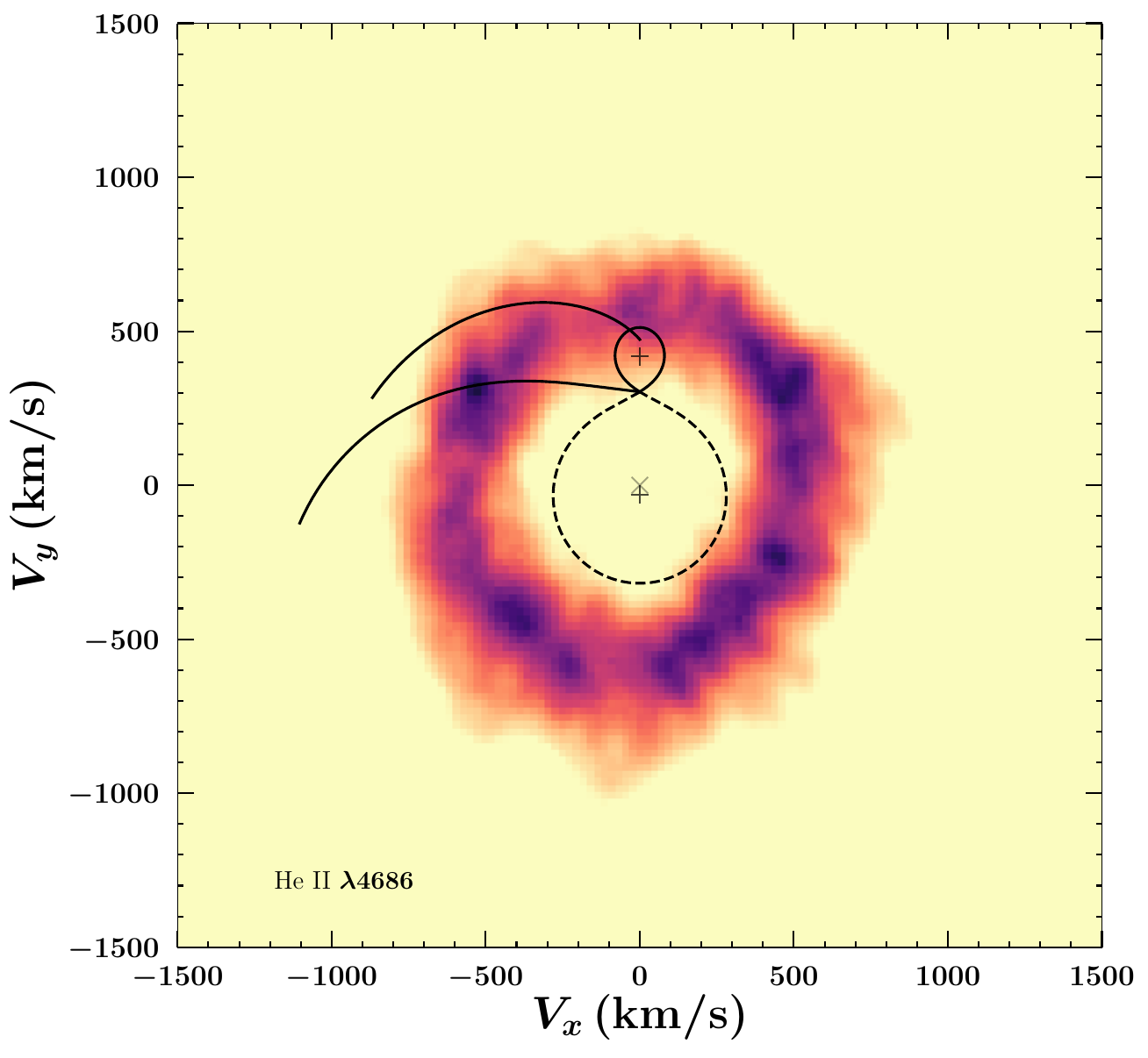}
    \includegraphics[width=0.34\linewidth,height=0.31\linewidth]{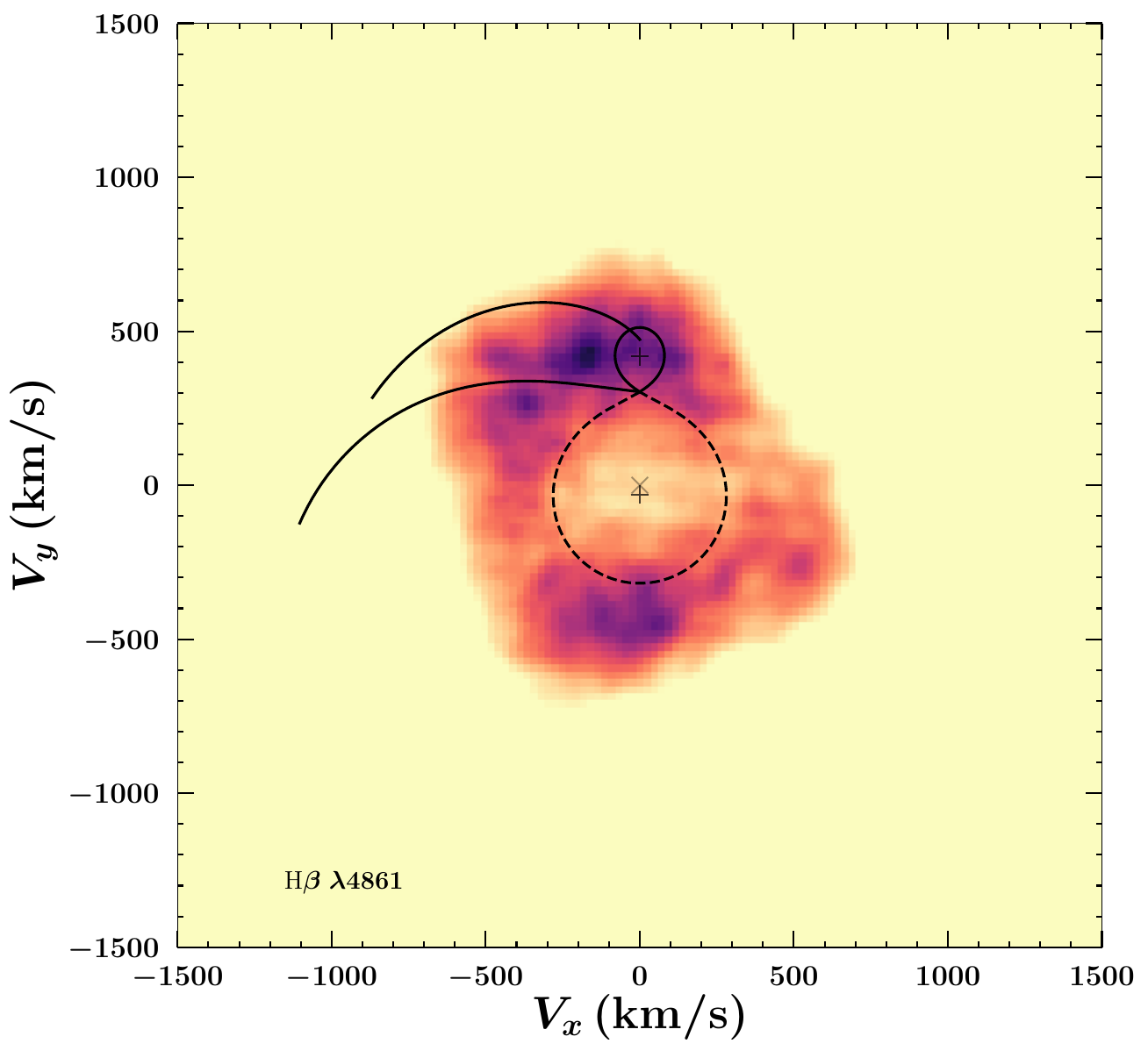}
    \includegraphics[width=0.34\linewidth,height=0.31\linewidth]{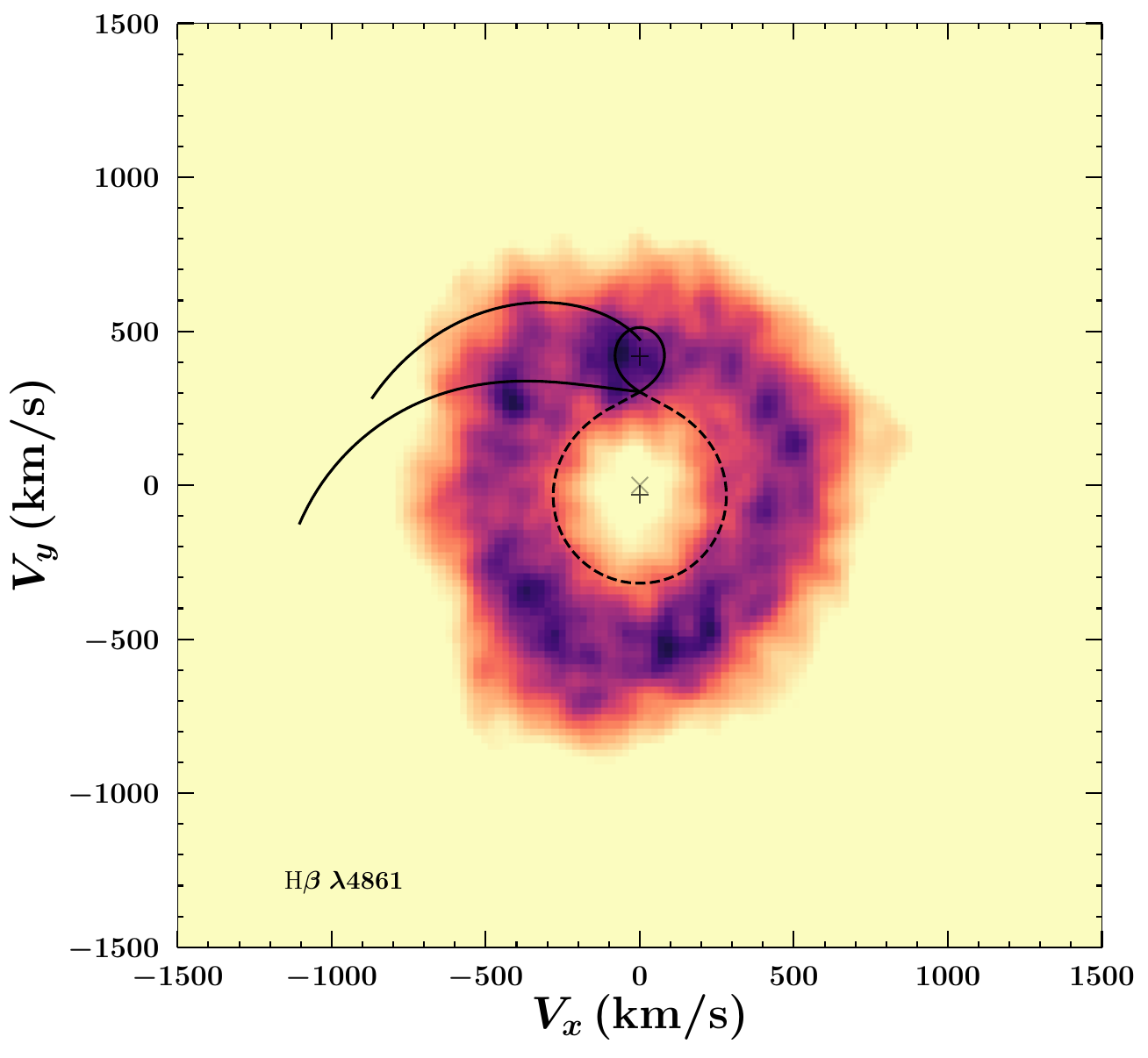}
    \includegraphics[width=0.34\linewidth,height=0.31\linewidth]{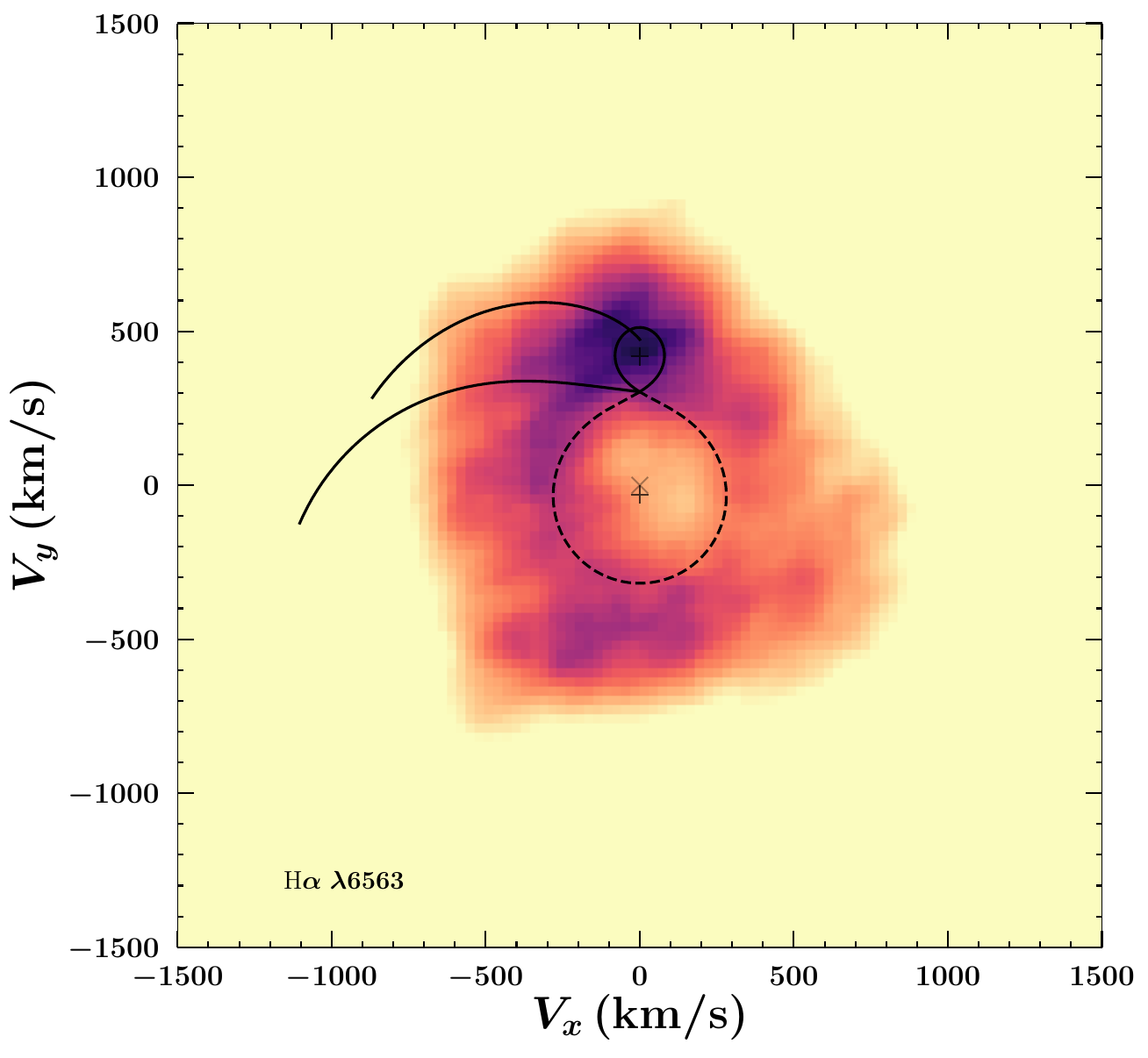}
    \includegraphics[width=0.34\linewidth,height=0.31\linewidth]{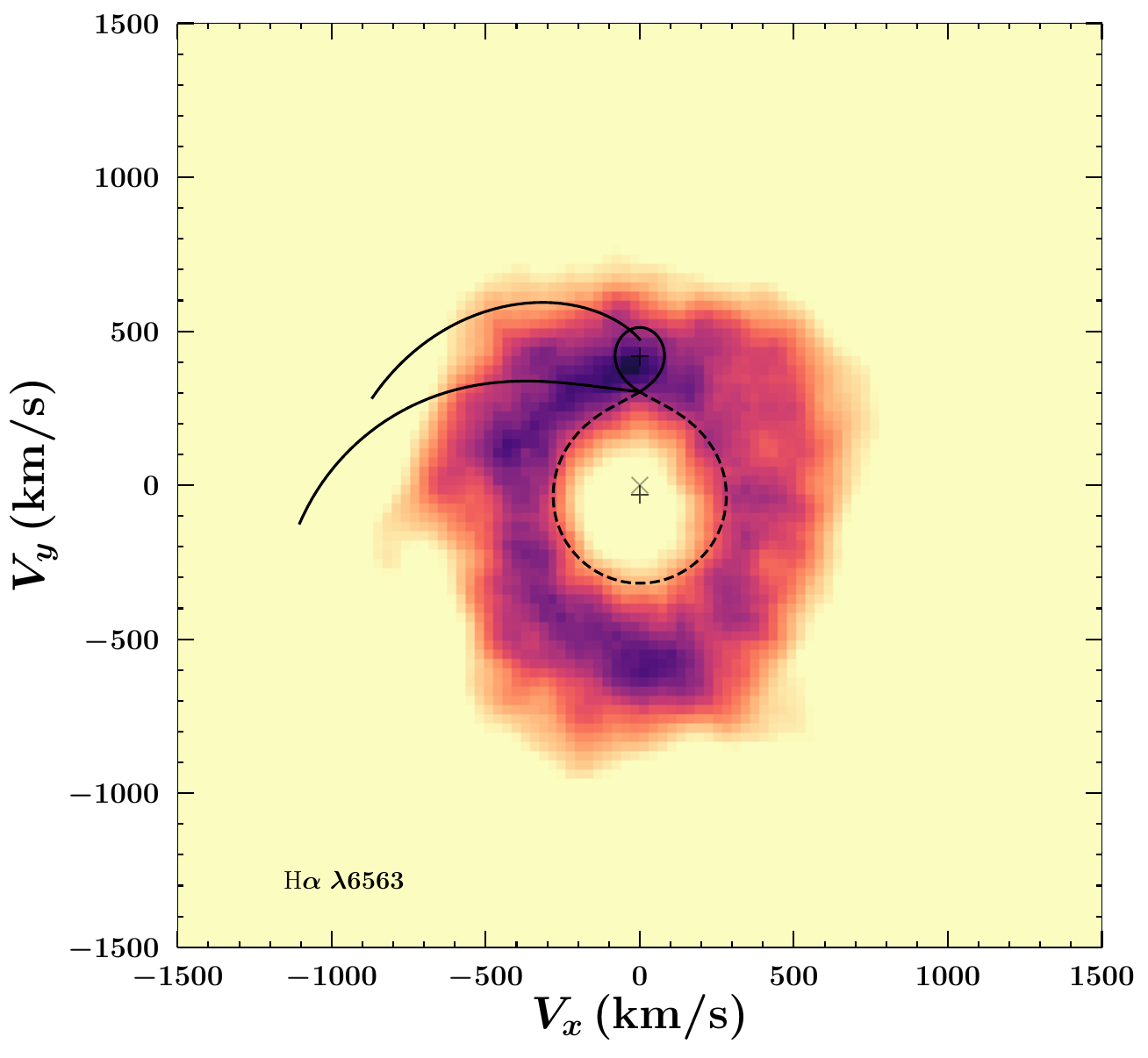}
    \includegraphics[width=0.34\linewidth,height=0.31\linewidth]{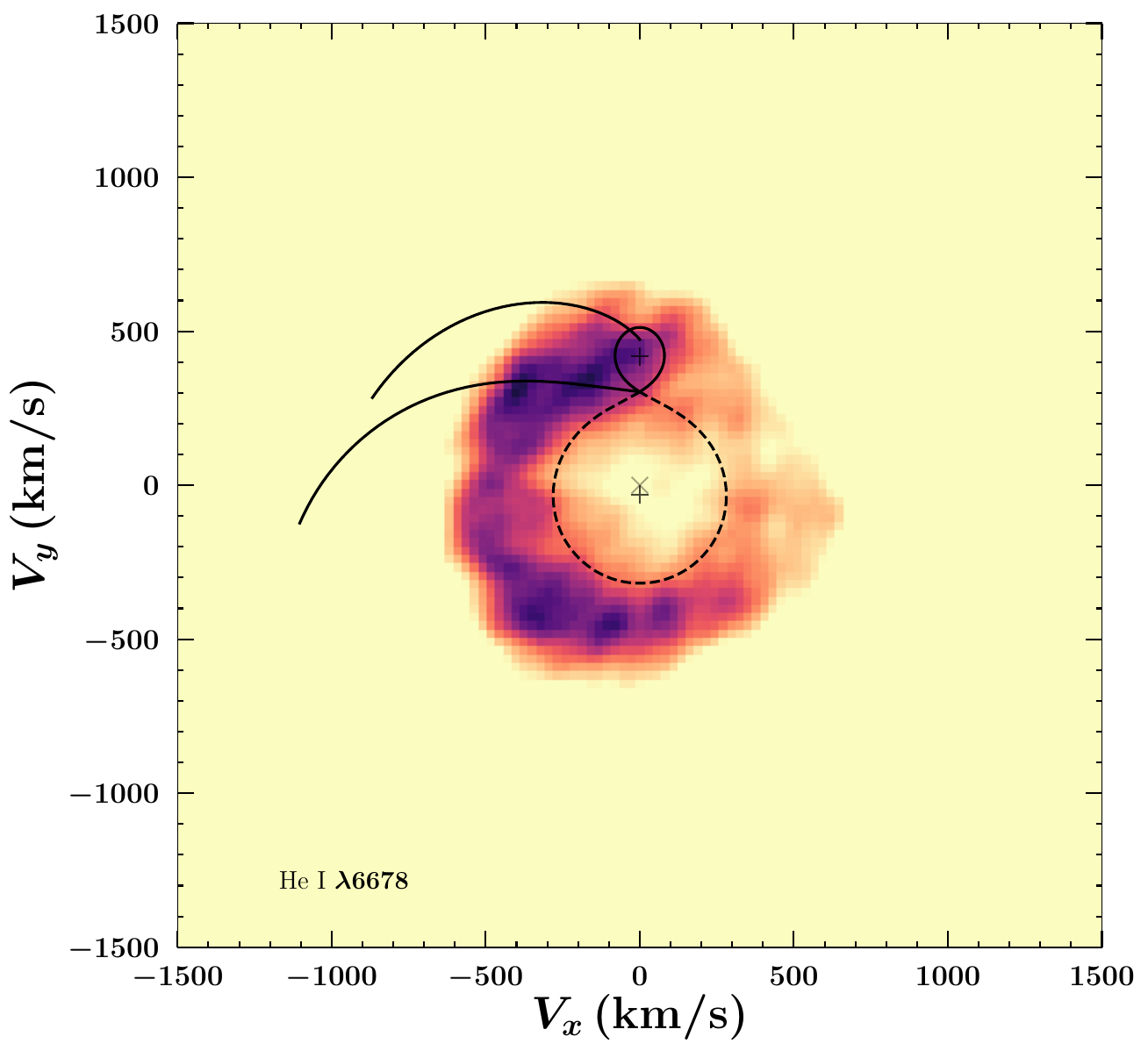}
    \includegraphics[width=0.34\linewidth,height=0.31\linewidth]{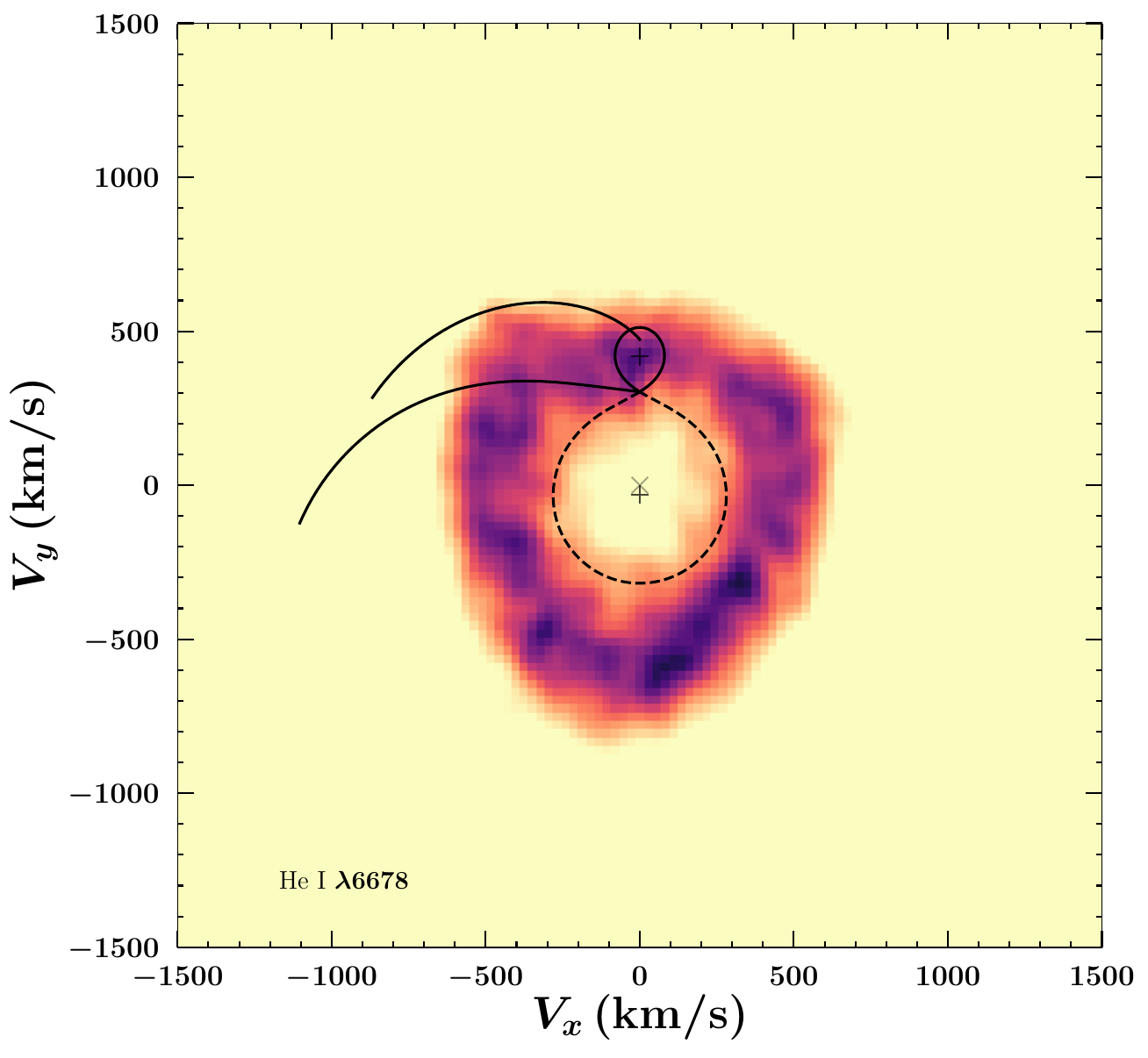}
    \caption{Doppler tomograms of (\textit{top to bottom}): He {\sc ii} $\lambda4686$, H$\beta$, H$\alpha$, and He {\sc i} $\lambda6678$ disc emission in the \textit{(left)} H/HIM and \textit{(right)} S/SIM accretion states. Over-plotted are the Roche lobes of the compact object (dashed line) and companion star (solid line) using J1820s known orbital parameters. Tomograms are created using all available GTC/OSIRIS and LT/FRODOspec data. See Section \ref{sec:dp_stuff} for details.}
    \label{fig:dp1}
\end{figure*}

\begin{figure*}
    \center

    \includegraphics[width=0.34\linewidth,height=0.31\linewidth]{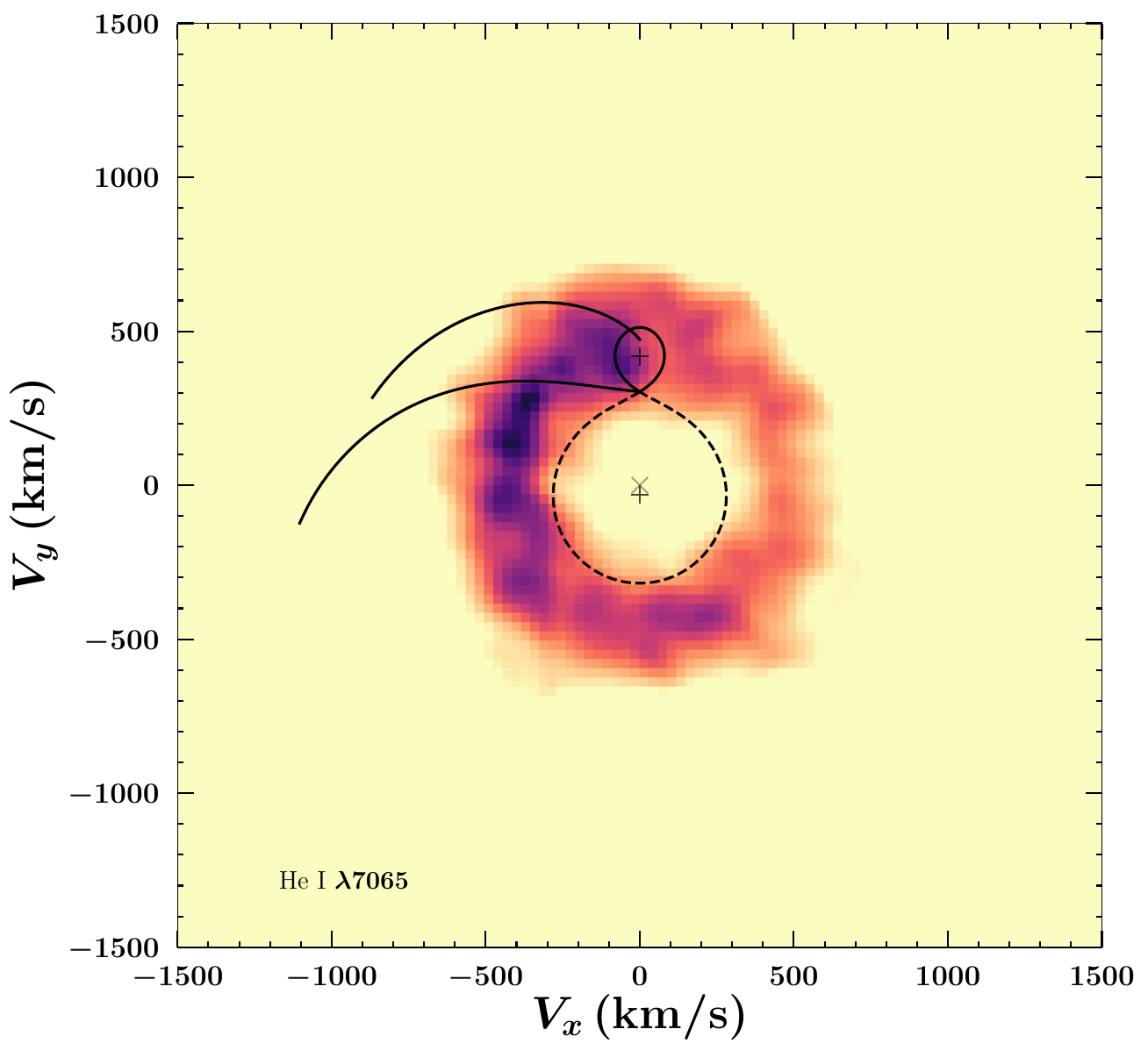}
    \includegraphics[width=0.34\linewidth,height=0.31\linewidth]{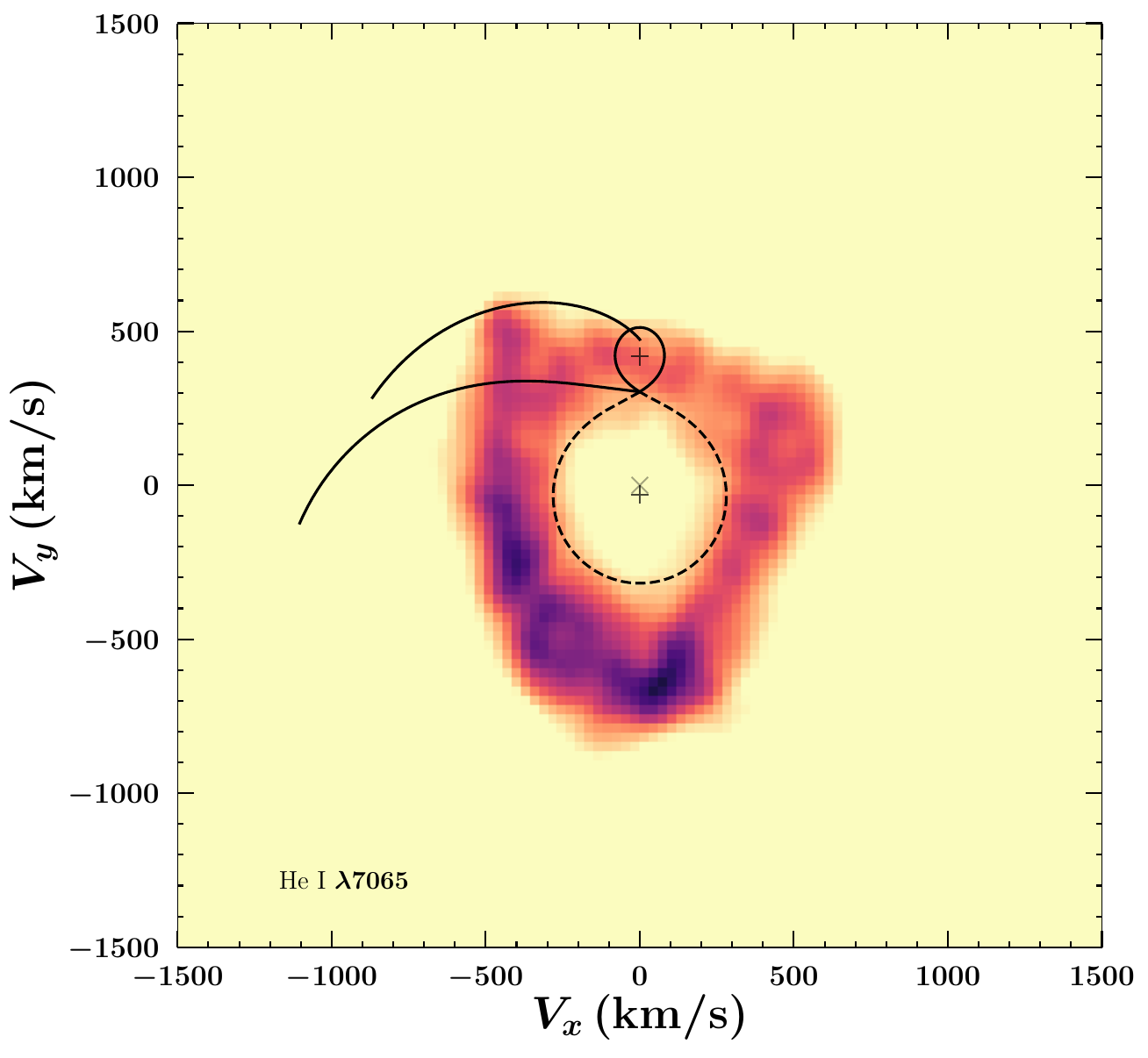}
    \caption{Same as figure \ref{fig:dp1} but for He {\sc i} $\lambda7065$.
    }
    \label{fig:dp2}
\end{figure*}

\subsection{Doppler Tomography}\label{sec:dp_stuff}

To date, Doppler tomography has been used to analyze complex emission line profiles from multitudes of LMXBs during outburst and quiescence (e.g., \citealt{casares1995,steeghs2004,davanzo2005,gonzalezhernandez2010,shaw2016,tetarenko2021,killestein2023}). However, it is important to note that, interpreting tomograms made during outburst is not a simple task. Doppler tomography assumes emission line flux does not vary over the binary orbit. When violated (during outburst), this can result in artifacts in the tomograms, making interpretation difficult. To combat this problem, data coverage over multiple binary orbits is necessary. An average across enough orbital cycles will overcome short-term variations and allow for clear interpretation of the outburst tomograms.
Considering the complete spectroscopic phase coverage available, and long duration, of the 2018 outburst of J1820, this system has presented us with the best opportunity for performing tomographic studies of outbursting LMXBs that we have had in decades.

We have created Doppler tomograms\footnote{Tom Marsh's {\sc molly} and {\tt doppler} (\url{https://github.com/trmrsh/trm-doppler}) software packages are used.} of the H$\alpha$, H$\beta$, He {\sc ii} $\lambda4686$, He {\sc i} $\lambda6678$, and He {\sc i} $\lambda7065$ emission from J1820 using a combination of the GTC and LT data-sets (see Sections \ref{sec:LT_data} and \ref{sec:GTC_data}). We split the data-sets into accretion state creating two tomograms for each line, one for the H/HIM, and one for the S/SIM, state data (see Figures \ref{fig:dp1} and \ref{fig:dp2}).
Tomograms were created using the known ephemeris of J1820 from \citet{torres2019}, and the following orbital parameters: $P_{\rm orb}=0.68549\pm0.00001$d, $q=0.072\pm0.012$, $M_1=8.48^{+0.79}_{-0.72}M_{\odot}$, and inclination $i=63^{\circ}\pm3$ \citep{torres2019,torres2020}.

\begin{table*}
	\centering
	\caption{Linear MCMC Fits to Empirical Correlations in J1820}
	\medskip
	\label{tab:corr_fits}
	\begin{tabular}{lcccccccc} 
		\hline
                & \multicolumn{4}{c}{EW vs $\mathcal{C}$}&\multicolumn{4}{c}{FWHM vs $\mathcal{C}$}\\\hline
		Emission&\multicolumn{2}{c}{\underline{ Hard State Fit}}&\multicolumn{2}{c}{\underline{ Soft State Fit}}&\multicolumn{2}{c}{\underline{ Hard State Fit}}&\multicolumn{2}{c}{\underline{ Soft State Fit}} \\
		Line&$m$&$b$&$m$&$b$&$m$&$b$&$m$&$b$\\
		\hline
He {\sc ii} $\lambda4686$ \AA & $-2.72_{-0.25}^{+0.28}$ & $16.06_{-0.87}^{+0.81}$ & $0.26\pm0.09$ & $-0.39_{-0.51}^{+0.56}$& $0.0086_{-0.0012}^{+0.0014}$ &$-3.99_{-1.60}^{+1.40}$&$0.0029_{-0.0010}^{+0.0011}$ &$-2.98_{-1.54}^{+1.45}$\\[0.05cm]
H$\beta$ $\lambda4861$ \AA & $-4.83_{-0.66}^{+0.58}$ &$15.11_{-0.88}^{+0.98}$&$0.28_{-0.18}^{+0.16}$& $0.46_{-0.39}^{+0.48}$&$0.0075_{-0.0017}^{+0.0019}$ &$0.19_{-1.89}^{+1.74}$&$0.0015_{-0.00061}^{+0.00064}$& $-0.76_{-0.85}^{+0.80}$ \\[0.05cm]
H$\alpha$ $\lambda6563$ \AA & $-1.11_{-0.16}^{+0.17}$ &$14.20_{-1.11}^{+1.01}$&$0.013_{-0.066}^{+0.063}$ &$1.12_{-0.41}^{+0.44}$
&$0.019_{-0.0023}^{+0.0029}$& $-16.01_{-3.45}^{+3.23}$&$0.0016_{-0.00074}^{+0.00079}$& $-0.60_{-0.92}^{+0.85}$ \\[0.05cm]
He {\sc i} $\lambda6678$ \AA & $-11.58_{-1.67}^{+1.27}$& $17.31_{-1.24}^{+1.01}$&
$1.69_{-0.58}^{+1.11}$& $-0.89_{-1.34}^{+0.72}$&$0.011_{-0.0017}^{+0.0018}$& $-2.00_{-1.41}^{+1.32}$&$0.0011_{-0.00053}^{+0.00055}$& $-0.0014_{-0.60}^{+0.58}$ \\[0.05cm]
He {\sc i} $\lambda7065$ \AA & $-6.71_{-0.85}^{+0.82}$ &$14.95_{-0.92}^{+0.88}$&
$1.34_{-0.43}^{+0.58}$ & $0.23_{-0.44}^{+0.34}$&
$0.021_{-0.0030}^{+0.0031}$& $-7.59_{-2.20}^{+2.00}$&
$0.00075_{-0.00054}^{+0.00054}$& $0.66_{-0.44}^{+0.43}$\\[0.05cm]
		\hline
\multicolumn{9}{p{2\columnwidth}}{\hangindent=1ex $^{*}$Each correlation was fit with a standard linear model: $y=mx+b$, where $m$ is the slope and $b$ is the y-intercept (see Section \ref{sec:correlations}).} \\
	\end{tabular}
\end{table*}

\section{Discussion}

\subsection{Hard State Correlations}

For all five lines of interest (H$\alpha$, H$\beta$, He {\sc ii} $\lambda4686$, He {\sc i} $\lambda6678$, and He {\sc i} $\lambda7065$), during the hard state we observe an increase in the reprocessed X-ray fraction ($\mathcal{C}$) correspond to an increase in FWHM, and a decrease in EW. We hypothesize that this behaviour is the observational signature for the presence, or at least onset, of a warped outer disc. A warp would cause the available reprocessing area in the outer disc to increase substantially (resulting in an increase in EW). As we observe an increase in EW to coincide with a decrease in FWHM, this tells us that the optical flux is dominated by material further out in the disc, where velocities are lower (i.e., the brightest part of the warp must shift to further out in the disc as the source evolves through the hard state). 

The behaviour seen in our correlations fit well with the scenario posited by \citet{thomasJ2022}, where a change in inner disc geometry occurs, allowing the source of hard X-ray irradiation, emitted close to the BH, to reach heights that allow for it to illuminate more and more of the warped outer disc region during the hard state. 
Early in the outburst, we observe large variations in FWHM (best seen with He {\sc ii} $\lambda4686$ and H$\beta$, as they probe much more of the inner disc regions than the H$\alpha$ and He {\sc i} lines), as the high hard X-ray flux begins to illuminate larger disc radii. Eventually, EW reaches its largest value (and FWHM is at its smallest) at MJD$58275$ (2018 June 6; (MJD$-58188)=87$), which coincides with when high amplitude modulations (``superhumps'') are first observed in the optical outburst light-curves. From MJD$58275$ (day 87) onwards, the evolution of $\mathcal{C}$ stops declining and remains high until the transition to the soft state begins, which is expected if the X-ray irradiation is in fact driving the warp, as postulated by \citet{thomasJ2022}.

Interestingly, Doppler tomograms made during the H/HIM state (see left panel of Figures \ref{fig:dp1} and \ref{fig:dp2}) show asymmetric patches of emission, in line with this hypothesis. The warp ``turns on'' (as evidenced by the observation of large optical modulations in the light-curve) at MJD$58275$ (2018 June 6; (MJD$-58188)=87$), covering the last $\sim30$d of the hard state. At this time, the outer disc structure must be evolving in parallel with the inner disc region filling up, as the source approaches the state transition. Ultimately, this produces asymmetric patches in the tomograms, as a result of a combination of bright parts of the outer warped disc and bright emission from material at higher velocities (and smaller radii). 

With all this being said, we note that there exists alternative explanations that could possibly explain our observations (e.g., hot spot emission, heated face of the donor star, X-ray modulation). However, as these possibilities have been strongly disfavoured by the independent analysis presented in \citet{thomasJ2022}, we favour the warped disc interpretation discussed here.

\subsection{Soft State Correlations}

For all five lines of interest 
during the soft state, we observe very different behaviour. In He {\sc ii} $\lambda4686$ and H$\beta$, we find that both FWHM and EW are tightly positively correlated with $\mathcal{C}$. 
Here FWHM and EW increase throughout the soft state. This tells us that the area of the disc illuminated by X-ray irradiation must shift further in as the source evolves through the soft state. Doppler tomograms made during the soft state (particularly in He {\sc ii} $\lambda4686$, and H$\beta$; see right panel of Figure \ref{fig:dp1}) show a more symmetric, full-disc structure. This is expected if bright emission at smaller disc radii is dominant here (as a result of more of the inner disc being illuminated).

In the H$\alpha$ and He {\sc i} lines, substantial variation in EW, paired with a relatively ``flat'' evolution (hovering around 
$\mathcal{C} \sim10^{-4}$), is observed along with asymmetry in the Doppler tomograms (see inset axes in Figures \ref{fig:corrs1}, \ref{fig:corrs2} and right panel of Figures \ref{fig:dp1} and \ref{fig:dp2}). This behaviour is expected, given that modulations in the optical light-curve remain high during the soft state time period (see Figure 6 of \citealp{thomasJ2022}), and these lines best probe the behaviour in the outer regions of the disc associated with the warp.

\section{Summary}

Disc-formed optical emission lines, formed as a result of high energy X-rays (emitted close to the BH) illuminating the outer disc, are effective observational tracers of how matter in LMXB accretion discs behaves and evolves throughout an outburst cycle. The shape, profile, and appearance/disappearance of these lines change throughout a binary orbit and carry the imprint of the evolving X-ray irradiation source heating the accretion disc throughout outburst. In this paper, we present a method to quantify this empirical connection existing between the line emitting regions, and physical properties of the X-ray source heating the disc, in LMXB systems, and demonstrate how these empirical correlations can be used as an effective observational tool towards understanding the structure and geometry of the gas making up an outbursting LMXB accretion disc.

Using a combination of phase-resolved optical spectroscopy from GTC and the Liverpool Telescopes, paired with X-ray, optical, and UV monitoring with \textit{Swift}, we show how changes in emission line profile shape (defined by FWHM and EW) are correlated with the evolving strength of high energy X-rays illuminating the disc during the 2018 outburst of BH-LMXB J1820. Focusing on the five strongest lines (H$\alpha$, H$\beta$, He {\sc ii} $\lambda4686$, He {\sc i} $\lambda6678$, and He {\sc i} $\lambda7065$) in the optical spectra, we are able to probe multiple different regions of the disc. Splitting the correlations into accretion state, creating state specific Doppler tomograms, and comparing hard vs soft state behaviour, has allowed us to track and quantify how variations in the X-ray irradiation heating during the outburst cycle affect physical properties of the gas in the accretion disc. Ultimately, we are able to confirm the scenario initially posited using comphrehensive timing analysis (e.g., \citealt{thomasJ2022}), for the presence of an irradiation-driven warped outer accretion disc in the system.

\section*{Acknowledgements}
BET acknowledges support from the Trottier Space Institute at McGill, through an MSI Fellowship and NSERC Discovery Grant RGPAS-2021-00021. BET is thankful to T. Mu\~{n}oz-Darias for providing the GTC data-set.
We greatly acknowledge the use of {\sc molly} and {\tt doppler} software packages developed by the late Tom Marsh, and thank J.V. Hernandez Santisteban for use of his {\sc python} scripts to plot Doppler tomograms.
This research is based on observations made with the: (i) LT operated on the island of La Palma by Liverpool John Moores University in the Spanish Observatorio del Roque de los Muchachos (ORM) of the Instituto de Astrofisica de Canarias with financial support from the UK Science and Technology Facilities Council, and (ii) GTC, also installed at the ORM. This research has also made use of: (i) data, software, and/or web tools obtained from the High Energy Astrophysics Science Archive Research Center (HEASARC), a service of the Astrophysics Science Division at NASA Goddard Space Flight Center (GSFC) and of the Smithsonian Astrophysical Observatory’s High Energy Astrophysics Division, (ii) data supplied by the UK Swift Science Data Centre at the University of Leicester, and (iii) NASA's Astrophysics Data System (ADS).

\section*{Data Availability}

 The observational data presented in this work are available online in the: HEASARC Archive (\url{https://heasarc.gsfc.nasa.gov/docs/archive.html}), LT Archive (Proposal IDs: PL17A10 - PI: A.W. Shaw, CL18A05 - PI: T. Mu\~{n}oz-Darias; \url{https://telescope.livjm.ac.uk/cgi-bin/lt_search}), and GTC Public Archive (Program IDs: GTC39-18A and GTC49-18B - PI: T. Mu\~{n}oz-Darias; \url{https://gtc.sdc.cab.inta-csic.es/gtc/jsp/searchform.jsp}).



\bibliographystyle{mnras}
\bibliography{paper_refs} 

\begin{thebibliography}{}
\makeatletter
\relax
\def\mn@urlcharsother{\let\do\@makeother \do\$\do\&\do\#\do\^\do\_\do\%\do\~}
\def\mn@doi{\begingroup\mn@urlcharsother \@ifnextchar [ {\mn@doi@}
  {\mn@doi@[]}}
\def\mn@doi@[#1]#2{\def\@tempa{#1}\ifx\@tempa\@empty \href
  {http://dx.doi.org/#2} {doi:#2}\else \href {http://dx.doi.org/#2} {#1}\fi
  \endgroup}
\def\mn@eprint#1#2{\mn@eprint@#1:#2::\@nil}
\def\mn@eprint@arXiv#1{\href {http://arxiv.org/abs/#1} {{\tt arXiv:#1}}}
\def\mn@eprint@dblp#1{\href {http://dblp.uni-trier.de/rec/bibtex/#1.xml}
  {dblp:#1}}
\def\mn@eprint@#1:#2:#3:#4\@nil{\def\@tempa {#1}\def\@tempb {#2}\def\@tempc
  {#3}\ifx \@tempc \@empty \let \@tempc \@tempb \let \@tempb \@tempa \fi \ifx
  \@tempb \@empty \def\@tempb {arXiv}\fi \@ifundefined
  {mn@eprint@\@tempb}{\@tempb:\@tempc}{\expandafter \expandafter \csname
  mn@eprint@\@tempb\endcsname \expandafter{\@tempc}}}

\bibitem[\protect\citeauthoryear{{Atri} et~al.,}{{Atri}
  et~al.}{2020}]{atri2020}
{Atri} P.,  et~al., 2020, \mn@doi [\mnras] {10.1093/mnrasl/slaa010}, \href
  {https://ui.adsabs.harvard.edu/abs/2020MNRAS.493L..81A} {493, L81}

\bibitem[\protect\citeauthoryear{{Baglio}, {Russell}  \& {Lewis}}{{Baglio}
  et~al.}{2018}]{baglio2018}
{Baglio} M.~C.,  {Russell} D.~M.,   {Lewis} F.,  2018, The Astronomer's
  Telegram, \href {https://ui.adsabs.harvard.edu/abs/2018ATel11418....1B}
  {11418, 1}

\bibitem[\protect\citeauthoryear{{Barnsley}, {Smith}  \& {Steele}}{{Barnsley}
  et~al.}{2012}]{barnsley2012}
{Barnsley} R.,  {Smith} R.,   {Steele} I.,  2012, in {Ballester} P.,  {Egret}
  D.,   {Lorente} N.~P.~F.,  eds,  Astronomical Society of the Pacific
  Conference Series Vol. 461, Astronomical Data Analysis Software and Systems
  XXI. p.~517

\bibitem[\protect\citeauthoryear{{Bright}, {Fender}  \& {Motta}}{{Bright}
  et~al.}{2018}]{bright2018}
{Bright} J.,  {Fender} R.,   {Motta} S.,  2018, The Astronomer's Telegram,
  \href {https://ui.adsabs.harvard.edu/abs/2018ATel11420....1B} {11420, 1}

\bibitem[\protect\citeauthoryear{{Bright} et~al.,}{{Bright}
  et~al.}{2020}]{bright2020}
{Bright} J.~S.,  et~al., 2020, \mn@doi [Nature Astronomy]
  {10.1038/s41550-020-1023-5}, \href
  {https://ui.adsabs.harvard.edu/abs/2020NatAs...4..697B} {4, 697}

\bibitem[\protect\citeauthoryear{{Burrows} et~al.,}{{Burrows}
  et~al.}{2005}]{burrows2005}
{Burrows} D.~N.,  et~al., 2005, \mn@doi [\ssr] {10.1007/s11214-005-5097-2},
  \href {https://ui.adsabs.harvard.edu/abs/2005SSRv..120..165B} {120, 165}

\bibitem[\protect\citeauthoryear{{Casares}}{{Casares}}{2015}]{casares2015}
{Casares} J.,  2015, \mn@doi [ApJ] {10.1088/0004-637X/808/1/80}, \href
  {https://ui.adsabs.harvard.edu/abs/2015ApJ...808...80C} {808, 80}

\bibitem[\protect\citeauthoryear{{Casares}, {Marsh}, {Charles}, {Martin},
  {Martin}, {Harlaftis}, {Pavlenko}  \& {Wagner}}{{Casares}
  et~al.}{1995}]{casares1995}
{Casares} J.,  {Marsh} T.~R.,  {Charles} P.~A.,  {Martin} A.~C.,  {Martin}
  E.~L.,  {Harlaftis} E.~T.,  {Pavlenko} E.~P.,   {Wagner} R.~M.,  1995,
  \mn@doi [MNRAS] {10.1093/mnras/274.2.565}, \href
  {https://ui.adsabs.harvard.edu/abs/1995MNRAS.274..565C} {274, 565}

\bibitem[\protect\citeauthoryear{{Cepa} et~al.,}{{Cepa}
  et~al.}{2000}]{cepa2000}
{Cepa} J.,  et~al., 2000, in {Iye} M.,  {Moorwood} A.~F.,  eds,  Society of
  Photo-Optical Instrumentation Engineers (SPIE) Conference Series Vol. 4008,
  Optical and IR Telescope Instrumentation and Detectors. pp 623--631,
  \mn@doi{10.1117/12.395520}

\bibitem[\protect\citeauthoryear{{Charles} \& {Coe}}{{Charles} \&
  {Coe}}{2006}]{charlescoe2006}
{Charles} P.~A.,  {Coe} M.~J.,  2006

\bibitem[\protect\citeauthoryear{{Crawford} \& {Kraft}}{{Crawford} \&
  {Kraft}}{1956}]{crawfordkraft1956}
{Crawford} J.~A.,  {Kraft} R.~P.,  1956, \mn@doi [ApJ] {10.1086/146128}, \href
  {https://ui.adsabs.harvard.edu/abs/1956ApJ...123...44C} {123, 44}

\bibitem[\protect\citeauthoryear{{D'Avanzo}, {Campana}, {Casares}, {Israel},
  {Covino}, {Charles}  \& {Stella}}{{D'Avanzo} et~al.}{2005}]{davanzo2005}
{D'Avanzo} P.,  {Campana} S.,  {Casares} J.,  {Israel} G.~L.,  {Covino} S.,
  {Charles} P.~A.,   {Stella} L.,  2005, \mn@doi [\aap]
  {10.1051/0004-6361:20053517}, \href
  {https://ui.adsabs.harvard.edu/abs/2005A&A...444..905D} {444, 905}

\bibitem[\protect\citeauthoryear{{Fabian} et~al.,}{{Fabian}
  et~al.}{2020}]{fabian2020}
{Fabian} A.~C.,  et~al., 2020, \mn@doi [\mnras] {10.1093/mnras/staa564}, \href
  {https://ui.adsabs.harvard.edu/abs/2020MNRAS.493.5389F} {493, 5389}

\bibitem[\protect\citeauthoryear{{Fitzpatrick} \& {Massa}}{{Fitzpatrick} \&
  {Massa}}{1999}]{fitzpatrick1999}
{Fitzpatrick} E.~L.,  {Massa} D.,  1999, \mn@doi [ApJ] {10.1086/307944}, \href
  {https://ui.adsabs.harvard.edu/abs/1999ApJ...525.1011F} {525, 1011}

\bibitem[\protect\citeauthoryear{{Foreman-Mackey}, {Hogg}, {Lang}  \&
  {Goodman}}{{Foreman-Mackey} et~al.}{2013}]{foremanmackay2013}
{Foreman-Mackey} D.,  {Hogg} D.~W.,  {Lang} D.,   {Goodman} J.,  2013, \mn@doi
  [\pasp] {10.1086/670067}, \href
  {https://ui.adsabs.harvard.edu/abs/2013PASP..125..306F} {125, 306}

\bibitem[\protect\citeauthoryear{{Gonz{\'a}lez Hern{\'a}ndez} \&
  {Casares}}{{Gonz{\'a}lez Hern{\'a}ndez} \&
  {Casares}}{2010}]{gonzalezhernandez2010}
{Gonz{\'a}lez Hern{\'a}ndez} J.~I.,  {Casares} J.,  2010, \mn@doi [\aap]
  {10.1051/0004-6361/201014088}, \href
  {https://ui.adsabs.harvard.edu/abs/2010A&A...516A..58G} {516, A58}

\bibitem[\protect\citeauthoryear{{Homan} et~al.,}{{Homan}
  et~al.}{2020}]{homan2020}
{Homan} J.,  et~al., 2020, \mn@doi [\apjl] {10.3847/2041-8213/ab7932}, \href
  {https://ui.adsabs.harvard.edu/abs/2020ApJ...891L..29H} {891, L29}

\bibitem[\protect\citeauthoryear{{Hubeny}}{{Hubeny}}{1990}]{hubeny1990}
{Hubeny} I.,  1990, \mn@doi [ApJ] {10.1086/168501}, \href
  {https://ui.adsabs.harvard.edu/abs/1990ApJ...351..632H} {351, 632}

\bibitem[\protect\citeauthoryear{{Kawamuro} et~al.,}{{Kawamuro}
  et~al.}{2018}]{kawamaro2018}
{Kawamuro} T.,  et~al., 2018, The Astronomer's Telegram, \href
  {https://ui.adsabs.harvard.edu/abs/2018ATel11399....1K} {11399, 1}

\bibitem[\protect\citeauthoryear{{Kennea}, {Marshall}, {Page}, {Palmer},
  {Siegel}  \& {Neil Gehrels Swift Observatory Team}}{{Kennea}
  et~al.}{2018}]{kennea2018}
{Kennea} J.~A.,  {Marshall} F.~E.,  {Page} K.~L.,  {Palmer} D.~M.,  {Siegel}
  M.~H.,   {Neil Gehrels Swift Observatory Team} 2018, The Astronomer's
  Telegram, \href {https://ui.adsabs.harvard.edu/abs/2018ATel11403....1K}
  {11403, 1}

\bibitem[\protect\citeauthoryear{{Killestein}, {Mould}, {Steeghs}, {Casares},
  {Galloway}  \& {Whelan}}{{Killestein} et~al.}{2023}]{killestein2023}
{Killestein} T.~L.,  {Mould} M.,  {Steeghs} D.,  {Casares} J.,  {Galloway}
  D.~K.,   {Whelan} J.~T.,  2023, \mn@doi [\mnras] {10.1093/mnras/stad366},
  \href {https://ui.adsabs.harvard.edu/abs/2023MNRAS.520.5317K} {520, 5317}

\bibitem[\protect\citeauthoryear{{Kim}, {Geem}  \& {Kim}}{{Kim}
  et~al.}{2001}]{geem2001}
{Kim} J.~H.,  {Geem} Z.~W.,   {Kim} E.~S.,  2001, \mn@doi [Journal of the
  American Water Resources Association] {10.1111/j.1752-1688.2001.tb03627.x},
  \href {https://ui.adsabs.harvard.edu/abs/2001JAWRA..37.1131K} {37, 1131}

\bibitem[\protect\citeauthoryear{{Marcel} et~al.,}{{Marcel}
  et~al.}{2019}]{marcel2019}
{Marcel} G.,  et~al., 2019, \mn@doi [A\&A] {10.1051/0004-6361/201935060}, \href
  {https://ui.adsabs.harvard.edu/abs/2019A&A...626A.115M} {626, A115}

\bibitem[\protect\citeauthoryear{{Marsh}}{{Marsh}}{2001}]{marsh2001}
{Marsh} T.~R.,  2001

\bibitem[\protect\citeauthoryear{{Marsh}}{{Marsh}}{2005}]{marsh2005}
{Marsh} T.~R.,  2005, \mn@doi [Ap\&SS] {10.1007/s10509-005-4859-3}, \href
  {https://ui.adsabs.harvard.edu/abs/2005Ap&SS.296..403M} {296, 403}

\bibitem[\protect\citeauthoryear{{Migliari} \& {Fender}}{{Migliari} \&
  {Fender}}{2006}]{migliari2006}
{Migliari} S.,  {Fender} R.~P.,  2006, \mn@doi [MNRAS]
  {10.1111/j.1365-2966.2005.09777.x}, \href
  {https://ui.adsabs.harvard.edu/abs/2006MNRAS.366...79M} {366, 79}

\bibitem[\protect\citeauthoryear{{Morales-Rueda}, {Carter}, {Steele}, {Charles}
   \& {Worswick}}{{Morales-Rueda} et~al.}{2004}]{MoralesRueda2004}
{Morales-Rueda} L.,  {Carter} D.,  {Steele} I.~A.,  {Charles} P.~A.,
  {Worswick} S.,  2004, \mn@doi [Astronomische Nachrichten]
  {10.1002/asna.200310228}, \href
  {https://ui.adsabs.harvard.edu/abs/2004AN....325..215M} {325, 215}

\bibitem[\protect\citeauthoryear{{Mu{\~n}oz-Darias} et~al.,}{{Mu{\~n}oz-Darias}
  et~al.}{2019}]{munozdarias2019}
{Mu{\~n}oz-Darias} T.,  et~al., 2019, \mn@doi [ApJL]
  {10.3847/2041-8213/ab2768}, \href
  {https://ui.adsabs.harvard.edu/abs/2019ApJ...879L...4M} {879, L4}

\bibitem[\protect\citeauthoryear{{Patterson} et~al.,}{{Patterson}
  et~al.}{2018}]{patterson2018}
{Patterson} J.,  et~al., 2018, The Astronomer's Telegram, \href
  {https://ui.adsabs.harvard.edu/abs/2018ATel11756....1P} {11756, 1}

\bibitem[\protect\citeauthoryear{{Remillard} \& {McClintock}}{{Remillard} \&
  {McClintock}}{2006}]{remillardmclintock2006}
{Remillard} R.~A.,  {McClintock} J.~E.,  2006, \mn@doi [ARA\&A]
  {10.1146/annurev.astro.44.051905.092532}, \href
  {https://ui.adsabs.harvard.edu/abs/2006ARA&A..44...49R} {44, 49}

\bibitem[\protect\citeauthoryear{{Roming} et~al.,}{{Roming}
  et~al.}{2005}]{roming2005}
{Roming} P. W.~A.,  et~al., 2005, \mn@doi [\ssr] {10.1007/s11214-005-5095-4},
  \href {https://ui.adsabs.harvard.edu/abs/2005SSRv..120...95R} {120, 95}

\bibitem[\protect\citeauthoryear{{Russell} et~al.,}{{Russell}
  et~al.}{2019a}]{russell2019a}
{Russell} D.~M.,  et~al., 2019a, \mn@doi [Astronomische Nachrichten]
  {10.1002/asna.201913610}, \href
  {https://ui.adsabs.harvard.edu/abs/2019AN....340..278R} {340, 278}

\bibitem[\protect\citeauthoryear{{Russell}, {Baglio}  \& {Lewis}}{{Russell}
  et~al.}{2019b}]{russell2019b}
{Russell} D.~M.,  {Baglio} M.~C.,   {Lewis} F.,  2019b, The Astronomer's
  Telegram, \href {https://ui.adsabs.harvard.edu/abs/2019ATel12534....1R}
  {12534, 1}

\bibitem[\protect\citeauthoryear{{Shaviv} \& {Wehrse}}{{Shaviv} \&
  {Wehrse}}{1986}]{shaviv1986}
{Shaviv} G.,  {Wehrse} R.,  1986, A\&A, \href
  {https://ui.adsabs.harvard.edu/abs/1986A&A...159L...5S} {159, L5}

\bibitem[\protect\citeauthoryear{{Shaw}, {Charles}, {Casares}  \&
  {Hern{\'a}ndez Santisteban}}{{Shaw} et~al.}{2016}]{shaw2016}
{Shaw} A.~W.,  {Charles} P.~A.,  {Casares} J.,   {Hern{\'a}ndez Santisteban}
  J.~V.,  2016, \mn@doi [\mnras] {10.1093/mnras/stw2092}, \href
  {https://ui.adsabs.harvard.edu/abs/2016MNRAS.463.1314S} {463, 1314}

\bibitem[\protect\citeauthoryear{{Shidatsu} et~al.,}{{Shidatsu}
  et~al.}{2018}]{shidatsu2018}
{Shidatsu} M.,  et~al., 2018, \mn@doi [\apj] {10.3847/1538-4357/aae929}, \href
  {https://ui.adsabs.harvard.edu/abs/2018ApJ...868...54S} {868, 54}

\bibitem[\protect\citeauthoryear{{Shidatsu}, {Nakahira}, {Murata}, {Adachi},
  {Kawai}, {Ueda}  \& {Negoro}}{{Shidatsu} et~al.}{2019}]{shidatsu2019}
{Shidatsu} M.,  {Nakahira} S.,  {Murata} K.~L.,  {Adachi} R.,  {Kawai} N.,
  {Ueda} Y.,   {Negoro} H.,  2019, \mn@doi [\apj] {10.3847/1538-4357/ab09ff},
  \href {https://ui.adsabs.harvard.edu/abs/2019ApJ...874..183S} {874, 183}

\bibitem[\protect\citeauthoryear{{Steeghs}}{{Steeghs}}{2004}]{steeghs2004}
{Steeghs} D.,  2004, \mn@doi [AN] {10.1002/asna.200310221}, \href
  {https://ui.adsabs.harvard.edu/abs/2004AN....325..185S} {325, 185}

\bibitem[\protect\citeauthoryear{{Tetarenko}, {Sivakoff}, {Heinke}  \&
  {Gladstone}}{{Tetarenko} et~al.}{2016}]{tetarenko2016}
{Tetarenko} B.~E.,  {Sivakoff} G.~R.,  {Heinke} C.~O.,   {Gladstone} J.~C.,
  2016, \mn@doi [ApJS] {10.3847/0067-0049/222/2/15}, \href
  {https://ui.adsabs.harvard.edu/abs/2016ApJS..222...15T} {222, 15}

\bibitem[\protect\citeauthoryear{{Tetarenko}, {Dubus}, {Marcel}, {Done}  \&
  {Clavel}}{{Tetarenko} et~al.}{2020}]{tetarenko2020}
{Tetarenko} B.~E.,  {Dubus} G.,  {Marcel} G.,  {Done} C.,   {Clavel} M.,  2020,
  \mn@doi [MNRAS] {10.1093/mnras/staa1367}, \href
  {https://ui.adsabs.harvard.edu/abs/2020MNRAS.495.3666T} {495, 3666}

\bibitem[\protect\citeauthoryear{{Tetarenko}, {Shaw}, {Manrow}, {Charles},
  {Miller}, {Russell}  \& {Tetarenko}}{{Tetarenko}
  et~al.}{2021}]{tetarenko2021}
{Tetarenko} B.~E.,  {Shaw} A.~W.,  {Manrow} E.~R.,  {Charles} P.~A.,  {Miller}
  J.~M.,  {Russell} T.~D.,   {Tetarenko} A.~J.,  2021, \mn@doi [\mnras]
  {10.1093/mnras/staa3861}, \href
  {https://ui.adsabs.harvard.edu/abs/2021MNRAS.501.3406T} {501, 3406}

\bibitem[\protect\citeauthoryear{{Thomas}, {Charles}, {Buckley}, {Kotze},
  {Lasota}, {Potter}, {Steiner}  \& {Paice}}{{Thomas}
  et~al.}{2022}]{thomasJ2022}
{Thomas} J.~K.,  {Charles} P.~A.,  {Buckley} D. A.~H.,  {Kotze} M.~M.,
  {Lasota} J.-P.,  {Potter} S.~B.,  {Steiner} J.~F.,   {Paice} J.~A.,  2022,
  \mn@doi [\mnras] {10.1093/mnras/stab3033}, \href
  {https://ui.adsabs.harvard.edu/abs/2022MNRAS.509.1062T} {509, 1062}

\bibitem[\protect\citeauthoryear{{Torres}, {Casares}, {Jim{\'e}nez-Ibarra},
  {Mu{\~n}oz-Darias}, {Armas Padilla}, {Jonker}  \& {Heida}}{{Torres}
  et~al.}{2019}]{torres2019}
{Torres} M.~A.~P.,  {Casares} J.,  {Jim{\'e}nez-Ibarra} F.,  {Mu{\~n}oz-Darias}
  T.,  {Armas Padilla} M.,  {Jonker} P.~G.,   {Heida} M.,  2019, \mn@doi
  [\apjl] {10.3847/2041-8213/ab39df}, \href
  {https://ui.adsabs.harvard.edu/abs/2019ApJ...882L..21T} {882, L21}

\bibitem[\protect\citeauthoryear{{Torres}, {Casares}, {Jim{\'e}nez-Ibarra},
  {{\'A}lvarez-Hern{\'a}ndez}, {Mu{\~n}oz-Darias}, {Armas Padilla}, {Jonker}
  \& {Heida}}{{Torres} et~al.}{2020}]{torres2020}
{Torres} M.~A.~P.,  {Casares} J.,  {Jim{\'e}nez-Ibarra} F.,
  {{\'A}lvarez-Hern{\'a}ndez} A.,  {Mu{\~n}oz-Darias} T.,  {Armas Padilla} M.,
  {Jonker} P.~G.,   {Heida} M.,  2020, \mn@doi [\apjl]
  {10.3847/2041-8213/ab863a}, \href
  {https://ui.adsabs.harvard.edu/abs/2020ApJ...893L..37T} {893, L37}

\bibitem[\protect\citeauthoryear{{Tucker} et~al.,}{{Tucker}
  et~al.}{2018a}]{tucker2018}
{Tucker} M.~A.,  et~al., 2018a, \mn@doi [\apjl] {10.3847/2041-8213/aae88a},
  \href {https://ui.adsabs.harvard.edu/abs/2018ApJ...867L...9T} {867, L9}

\bibitem[\protect\citeauthoryear{{Tucker} et~al.,}{{Tucker}
  et~al.}{2018b}]{tucker2018b}
{Tucker} M.~A.,  et~al., 2018b, \mn@doi [\apjl] {10.3847/2041-8213/aae88a},
  \href {https://ui.adsabs.harvard.edu/abs/2018ApJ...867L...9T} {867, L9}

\bibitem[\protect\citeauthoryear{{Verner}, {Ferland}, {Korista}  \&
  {Yakovlev}}{{Verner} et~al.}{1996}]{verner1996}
{Verner} D.~A.,  {Ferland} G.~J.,  {Korista} K.~T.,   {Yakovlev} D.~G.,  1996,
  \mn@doi [ApJ] {10.1086/177435}, \href
  {https://ui.adsabs.harvard.edu/abs/1996ApJ...465..487V} {465, 487}

\bibitem[\protect\citeauthoryear{{Wilms}, {Allen}  \& {McCray}}{{Wilms}
  et~al.}{2000}]{wilms2000}
{Wilms} J.,  {Allen} A.,   {McCray} R.,  2000, \mn@doi [ApJ] {10.1086/317016},
  \href {https://ui.adsabs.harvard.edu/abs/2000ApJ...542..914W} {542, 914}

\bibitem[\protect\citeauthoryear{{van Dokkum}}{{van
  Dokkum}}{2001}]{lacosmicpaper}
{van Dokkum} P.~G.,  2001, \mn@doi [PASP] {10.1086/323894}, \href
  {https://ui.adsabs.harvard.edu/abs/2001PASP..113.1420V} {113, 1420}

\bibitem[\protect\citeauthoryear{{van Paradijs}}{{van
  Paradijs}}{1996}]{vanparadijs1996}
{van Paradijs} J.,  1996, \mn@doi [ApJL] {10.1086/310100}, \href
  {https://ui.adsabs.harvard.edu/abs/1996ApJ...464L.139V} {464, L139}

\bibitem[\protect\citeauthoryear{{van Paradijs} \& {McClintock}}{{van Paradijs}
  \& {McClintock}}{1994}]{vanparadijs1994}
{van Paradijs} J.,  {McClintock} J.~E.,  1994, A\&A, \href
  {https://ui.adsabs.harvard.edu/abs/1994A%26A...290..133V} {290, 133}

\makeatother
\end{thebibliography}




\appendix

\section{Optical Spectroscopy Observation Logs}

\begin{table}
	\centering
	\caption{GTC Observation Log: Grating R2500V}
	\medskip
	\label{tab:gtc_v}
	\begin{tabular}{ccc} 
		\hline
		Spectra&Exposure&UTC\\
            Number&Time (s)&\\
		\hline
1 & 200.0 & 17/03/18 05:30:38 \\
2 & 200.0 & 18/03/18 05:59:59\\
3 & 200.0 & 18/03/18 06:03:44\\
4 & 200.0 & 20/03/18 05:48:04\\
5 & 200.0 & 20/03/18 05:52:18\\
6 & 200.0 & 21/03/18 06:01:42\\
7 & 200.0 & 21/03/18 06:05:26\\
8 & 180.0 & 22/03/18 05:25:00\\
9 & 180.0 & 22/03/18 05:30:43\\
10 & 90.00 & 24/03/18 05:29:35\\
11 & 90.00 & 24/03/18 05:31:28\\
12 & 90.00 & 24/03/18 05:33:22\\
13 & 180.0 & 26/03/18 03:35:47\\
14 & 180.0 & 26/03/18 03:39:10\\
15 & 180.0 & 26/03/18 03:49:15\\
16 & 180.0 & 26/03/18 03:52:41\\
17 & 180.0 & 26/03/18 05:49:30\\
18 & 180.0 & 26/03/18 05:52:53\\
19 & 180.0 & 08/07/18 02:10:04\\
20 & 180.0 & 08/07/18 02:13:27\\
21 & 180.0 & 10/07/18 21:37:47\\
22 & 180.0 & 10/07/18 21:41:10\\
23 & 180.0 & 11/07/18 21:30:25\\
24 & 180.0 & 11/07/18 21:33:50\\
25 & 180.0 & 15/07/18 01:32:37\\
26 & 180.0 & 15/07/18 01:36:02\\
27 & 180.0 & 18/07/18 01:47:00\\
28 & 180.0 & 18/07/18 01:50:23\\
29 & 180.0 & 18/07/18 21:46:41\\
30 & 180.0 & 18/07/18 21:50:05\\
31 & 180.0 & 24/07/18 22:53:59\\
32 & 180.0 & 24/07/18 22:57:22\\
33 & 180.0 & 27/07/18 21:34:03\\
34 & 180.0 & 27/07/18 21:37:26\\
35 & 180.0 & 03/08/18 23:24:56\\
36 & 180.0 & 03/08/18 23:28:20\\
37 & 180.0 & 09/08/18 22:22:56\\
38 & 180.0 & 09/08/18 22:26:20\\
39 & 180.0 & 15/08/18 21:46:31\\
40 & 180.0 & 15/08/18 21:49:55\\
41 & 180.0 & 19/08/18 22:00:17\\
42 & 180.0 & 19/08/18 22:03:41\\
43 & 400.0 & 12/10/18 21:42:14\\
44 & 400.0 & 21/10/18 21:26:12\\
45 & 400.0 & 04/11/18 19:52:03\\
  		\hline
	\end{tabular}
 \label{tab:gtcV}
\end{table}

\begin{table}
	\centering
	\caption{GTC Observation Log: Grating R2500R}
	\medskip
	\label{tab:gtc_r}
	\begin{tabular}{ccc} 
		\hline
		Spectra&Exposure&UTC\\
            Number&Time (s)&\\
		\hline
  1&200.0&17/03/18 05:26:37\\
2&200.0&18/03/18 06:07:44\\
3&200.0&18/03/18 06:11:29\\
4&200.0&20/03/18 05:56:20\\
5&200.0&20/03/18 06:00:03\\
6&200.0&21/03/18 06:09:25\\
7&200.0&21/03/18 06:13:09\\
8&120.0&22/03/18 05:34:24\\
9&120.0&22/03/18 05:36:48\\
10&120.0&22/03/18 05:39:11\\
11&120.0&22/03/18 05:41:35\\
12&75.00&24/03/18 05:35:33\\
13&75.00&24/03/18 05:37:12\\
14&75.00&24/03/18 05:38:51\\
15&75.00&24/03/18 05:40:29\\
16&150.0&26/03/18 03:42:50\\
17&150.0&26/03/18 03:45:43\\
18&150.0&26/03/18 03:56:22\\
19&150.0&26/03/18 03:59:15\\
20&150.0&26/03/18 05:43:24\\
21&150.0&26/03/18 05:46:18\\
22&150.0&08/07/18 02:04:02\\
23&150.0&08/07/18 02:06:56\\
24&150.0&10/07/18 21:31:45\\
25&150.0&10/07/18 21:34:39\\
26&150.0&11/07/18 21:24:22\\
27&150.0&11/07/18 21:27:17\\
28&150.0&15/07/18 01:26:34\\
29&150.0&15/07/18 01:29:28\\
30&150.0&18/07/18 01:40:57\\
31&150.0&18/07/18 01:43:50\\
32&150.0&18/07/18 21:40:38\\
33&150.0&18/07/18 21:43:32\\
34&150.0&24/07/18 22:47:55\\
35&150.0&24/07/18 22:50:50\\
36&150.0&27/07/18 21:28:00\\
37&150.0&27/07/18 21:30:54\\
38&150.0&03/08/18 23:18:50\\
39&150.0&03/08/18 23:21:43\\
40&150.0&09/08/18 22:16:51\\
41&150.0&09/08/18 22:19:45\\
42&150.0&15/08/18 21:40:24\\
43&150.0&15/08/18 21:43:21\\
44&150.0&19/08/18 21:54:17\\
45&150.0&19/08/18 21:57:10\\
46&400.0&12/10/18 21:34:54\\
47&400.0&21/10/18 21:18:54\\
48&400.0&04/11/18 19:44:43\\
  		\hline
    \label{tab:gtcR}
	\end{tabular}
\end{table}

\clearpage
\captionsetup[table]{labelsep=period}
\tablecaption{FRODOspec Observation Log: Blue}	
\label{tab:frodoblue}
\tablefirsthead{%
    \hline
        Spectra & Exposure & UTC  \\
        Number & Time (s) &  \\\hline}
\begin{supertabular}{ccc}
    \hline
1 &  600.0  & 17/3/18 05:41:42.498\\
2 &  600.0  & 17/3/18 05:52:14.842\\
3 &  600.0  & 17/3/18 06:02:47.065\\
4 &  900.0  & 18/3/18 04:46:18.022\\
5 &  900.0  & 18/3/18 05:01:51.074\\
6 &  900.0  & 18/3/18 05:17:26.304\\
7 &  900.0  & 20/3/18 04:15:54.132\\
8 &  900.0  & 20/3/18 04:31:27.148\\
9 &  900.0  & 20/3/18 04:47:02.076\\
10 &  900.0  & 21/3/18 05:33:53.350\\
11 &  900.0  & 21/3/18 05:49:26.005\\
12 &  900.0  & 21/3/18 06:05:00.233\\
13 &  900.0  & 22/3/18 04:10:01.603\\
14 &  900.0  & 22/3/18 04:25:34.488\\
15 &  900.0  & 22/3/18 04:41:06.833\\
16 &  900.0  & 23/3/18 05:01:42.787\\
17 &  900.0  & 23/3/18 05:17:15.497\\
18 &  900.0  & 23/3/18 05:32:49.716\\
19 &  900.0  & 24/3/18 04:35:12.049\\
20 &  900.0  & 24/3/18 04:50:44.641\\
21 &  900.0  & 24/3/18 05:06:18.908\\
22 &  900.0  & 26/3/18 04:24:13.961\\
23 &  900.0  & 26/3/18 04:39:46.398\\
24 &  900.0  & 26/3/18 04:55:20.644\\
25 &  900.0  & 27/3/18 04:50:34.338\\
26 &  900.0  & 27/3/18 05:06:06.667\\
27 &  900.0  & 27/3/18 05:21:41.701\\
28 &  900.0  & 29/3/18 04:08:01.301\\
29 &  900.0  & 29/3/18 04:23:33.828\\
30 &  900.0  & 29/3/18 04:39:08.054\\
31 &  900.0  & 30/3/18 04:04:37.370\\
32 &  900.0  & 30/3/18 04:20:09.826\\
33 &  900.0  & 30/3/18 04:35:44.886\\
34 &  900.0  & 31/3/18 04:06:07.277\\
35 &  900.0  & 31/3/18 04:21:39.472\\
36 &  900.0  & 31/3/18 04:37:14.818\\
37 &  900.0  & 3/4/18 04:06:12.697\\
38 &  900.0  & 3/4/18 04:21:45.721\\
39 &  900.0  & 3/4/18 04:37:20.877\\
40 &  900.0  & 4/4/18 03:30:24.005\\
41 &  900.0  & 4/4/18 03:45:56.662\\
42 &  900.0  & 8/4/18 03:09:57.099\\
43 &  900.0  & 8/4/18 03:25:29.766\\
44 &  900.0  & 11/4/18 05:12:10.414\\
45 &  900.0  & 11/4/18 05:27:42.694\\
46 &  900.0  & 12/4/18 05:04:11.055\\
47 &  900.0  & 12/4/18 05:19:43.383\\
48 &  90.0  & 12/4/18 05:38:53.613\\
49 &  900.0  & 13/4/18 04:21:12.464\\
50 &  900.0  & 13/4/18 04:36:45.445\\
51 &  900.0  & 14/4/18 03:48:15.030\\
52 &  900.0  & 14/4/18 04:03:47.990\\
53 &  900.0  & 15/4/18 03:18:54.360\\
54 &  900.0  & 15/4/18 03:34:26.811\\
55 &  900.0  & 16/4/18 02:38:37.252\\
56 &  900.0  & 16/4/18 02:54:09.914\\
57 &  900.0  & 17/4/18 04:10:05.467\\
58 &  900.0  & 17/4/18 04:25:37.891\\
59 &  900.0  & 18/4/18 03:08:32.416\\
60 &  900.0  & 18/4/18 03:24:05.114\\
61 &  900.0  & 27/4/18 03:05:52.829\\
62 &  900.0  & 27/4/18 03:21:25.299\\
63 &  900.0  & 29/4/18 01:43:42.728\\
64 &  900.0  & 29/4/18 01:59:15.187\\
65 &  900.0  & 1/5/18 01:45:39.081\\
66 &  900.0  & 1/5/18 02:01:11.710\\
67 &  900.0  & 5/5/18 02:27:27.288\\
68 &  900.0  & 5/5/18 02:43:00.837\\
69 &  900.0  & 11/5/18 01:48:21.620\\
70 &  900.0  & 11/5/18 02:03:53.676\\
71 &  900.0  & 13/5/18 03:58:37.065\\
72 &  900.0  & 13/5/18 04:14:09.221\\
73 &  900.0  & 16/5/18 01:48:30.752\\
74 &  900.0  & 16/5/18 02:04:03.367\\
75 &  900.0  & 23/5/18 01:10:59.797\\
76 &  900.0  & 23/5/18 01:26:32.209\\
77 &  900.0  & 25/5/18 01:14:50.765\\
78 &  900.0  & 25/5/18 01:30:23.423\\
79 &  900.0  & 27/5/18 23:56:05.080\\
80 &  900.0  & 28/5/18 00:11:37.681\\
81 &  900.0  & 29/5/18 00:40:07.878\\
82 &  900.0  & 29/5/18 00:55:40.355\\
83 &  900.0  & 3/6/18 00:18:16.107\\
84 &  900.0  & 3/6/18 00:33:48.636\\
85 &  900.0  & 8/6/18 00:40:15.799\\
86 &  900.0  & 8/6/18 00:55:48.110\\
87 &  900.0  & 13/6/18 01:28:14.679\\
88 &  900.0  & 13/6/18 01:43:47.394\\
89 &  900.0  & 18/6/18 04:55:08.423\\
\hline

\end{supertabular}

\clearpage

\tablecaption{FRODOspec Observation Log: Red}	
\label{tab:frodored}
\tablefirsthead{%
    \hline
        Spectra & Exposure & UTC  \\
        Number & Time (s) &  \\\hline}
\begin{supertabular}{ccc}
    \hline
1 &  600.0  & 17/3/18 05:41:48.179\\
2 &  600.0  & 17/3/18 05:52:23.361\\
3 &  600.0  & 17/3/18 06:02:58.439\\
4 &  900.0  & 18/3/18 04:46:12.359\\
5 &  900.0  & 18/3/18 05:01:47.988\\
6 &  900.0  & 20/3/18 04:15:48.446\\
7 &  900.0  & 20/3/18 04:31:23.938\\
8 &  900.0  & 20/3/18 04:46:58.872\\
9 &  900.0  & 21/3/18 05:33:47.567\\
10 &  900.0  & 21/3/18 05:49:22.363\\
11 &  900.0  & 21/3/18 06:04:56.987\\
12 &  900.0  & 22/3/18 04:10:07.782\\
13 &  900.0  & 22/3/18 04:25:43.173\\
14 &  900.0  & 22/3/18 04:41:18.097\\
15 &  900.0  & 23/3/18 05:01:36.993\\
16 &  900.0  & 23/3/18 05:17:11.764\\
17 &  900.0  & 23/3/18 05:32:46.534\\
18 &  900.0  & 24/3/18 04:35:06.304\\
19 &  900.0  & 24/3/18 04:50:40.960\\
20 &  900.0  & 24/3/18 05:06:15.605\\
21 &  900.0  & 26/3/18 04:24:07.608\\
22 &  900.0  & 26/3/18 04:39:42.707\\
23 &  900.0  & 26/3/18 04:55:17.502\\
24 &  900.0  & 27/3/18 04:50:28.143\\
25 &  900.0  & 27/3/18 05:06:03.510\\
26 &  900.0  & 27/3/18 05:21:38.565\\
27 &  900.0  & 29/3/18 04:07:55.109\\
28 &  900.0  & 29/3/18 04:23:30.197\\
29 &  900.0  & 29/3/18 04:39:04.869\\
30 &  900.0  & 30/3/18 04:04:31.207\\
31 &  900.0  & 30/3/18 04:20:06.276\\
32 &  900.0  & 30/3/18 04:35:41.749\\
33 &  900.0  & 31/3/18 04:06:01.170\\
34 &  900.0  & 31/3/18 04:21:35.963\\
35 &  900.0  & 31/3/18 04:37:11.434\\
36 &  900.0  & 3/4/18 04:06:07.058\\
37 &  900.0  & 3/4/18 04:21:42.552\\
38 &  900.0  & 3/4/18 04:37:18.284\\
39 &  900.0  & 4/4/18 03:30:30.179\\
40 &  900.0  & 4/4/18 03:46:05.235\\
42 &  900.0  & 8/4/18 03:10:03.278\\
43 &  900.0  & 8/4/18 03:25:38.619\\
43 &  900.0  & 11/4/18 05:12:16.115\\
44 &  900.0  & 11/4/18 05:27:51.157\\
45 &  900.0  & 12/4/18 05:04:16.768\\
46 &  900.0  & 12/4/18 05:19:51.891\\
47 &  900.0  & 13/4/18 04:21:17.891\\
48 &  900.0  & 13/4/18 04:36:52.915\\
49 &  900.0  & 14/4/18 03:48:21.157\\
50 &  900.0  & 14/4/18 04:03:56.521\\
51 &  900.0  & 15/4/18 03:19:00.212\\
52 &  900.0  & 15/4/18 03:34:35.354\\
53 &  900.0  & 16/4/18 02:38:43.438\\
54 &  900.0  & 16/4/18 02:54:18.478\\
55 &  900.0  & 17/4/18 04:10:11.258\\
56 &  900.0  & 17/4/18 04:25:46.374\\
57 &  900.0  & 18/4/18 03:08:38.570\\
58 &  900.0  & 27/4/18 03:05:58.771\\
59 &  900.0  & 27/4/18 03:21:33.861\\
60 &  900.0  & 29/4/18 01:43:48.874\\
61 &  900.0  & 29/4/18 01:59:23.706\\
62 &  900.0  & 1/5/18 01:45:45.281\\
63 &  900.0  & 1/5/18 02:01:20.101\\
64 &  900.0  & 5/5/18 02:27:32.876\\
65 &  900.0  & 5/5/18 02:43:08.173\\
66 &  900.0  & 11/5/18 01:48:27.367\\
67 &  900.0  & 11/5/18 02:04:02.187\\
68 &  900.0  & 13/5/18 03:58:30.941\\
69 &  900.0  & 13/5/18 04:14:05.825\\
70 &  900.0  & 16/5/18 01:48:25.009\\
71 &  900.0  & 16/5/18 02:04:01.003\\
72 &  900.0  & 23/5/18 01:10:54.163\\
73 &  900.0  & 23/5/18 01:26:28.700\\
74 &  900.0  & 25/5/18 01:14:56.871\\
75 &  900.0  & 25/5/18 01:30:31.910\\
76 &  900.0  & 27/5/18 23:55:58.958\\
77 &  900.0  & 28/5/18 00:11:34.048\\
78 &  900.0  & 29/5/18 00:40:02.010\\
79 &  900.0  & 29/5/18 00:55:36.697\\
80 &  900.0  & 3/6/18 00:18:09.943\\
81 &  900.0  & 3/6/18 00:33:45.002\\
82 &  900.0  & 8/6/18 00:40:21.790\\
83 &  900.0  & 8/6/18 00:55:56.720\\
84 &  900.0  & 13/6/18 01:28:08.533\\
85 &  900.0  & 13/6/18 01:43:43.789\\
86 &  900.0  & 18/6/18 04:55:14.573\\
\hline

\end{supertabular}




\bsp	
\label{lastpage}
\end{document}